# Spatiotemporal wall pressure forecast of a rectangular cylinder with physics-aware DeepU-Fourier neural network

Junle Liu (刘军乐) ; Chang Liu (刘畅) ; Yanyu Ke (柯延宇) ; Wenliang Chen (陈文亮) ; Kihing Shum (沈其庆) ; Tim K. T. Tse (谢锦添) ; Gang Hu (胡钢)

# Spatiotemporal wall pressure forecast of a rectangular cylinder with physics-aware DeepU-Fourier neural network

Junle Liu (刘军乐),[1,2] Chang Liu (刘畅),[3] Yanyu Ke (柯延宇),[1] Wenliang Chen (陈文亮),[2] Kihing Shum (沈其庆),[1] Tim K. T. Tse (谢锦添),[1] and Gang Hu (胡钢)[2,a)]

## AFFILIATIONS

[1]Department of Civil and Environmental Engineering, The Hong Kong University of Science and Technology, Clear Water Bay, Kowloon, Hong Kong SAR, China
[2]Artificial Intelligence for Wind Engineering (AIWE) Lab, School of Intelligent Civil and Ocean Engineering, Harbin Institute of Technology, Shenzhen 518055, China
[3]School of Mechanical, Aerospace, and Manufacturing Engineering, University of Connecticut, Storrs, Connecticut 06269, USA

[a)]Author to whom correspondence should be addressed: hugang@hit.edu.cn

## ABSTRACT

The wall pressure is of great importance in understanding the forces and structural responses induced by fluid. Recent works have investigated the potential of deep learning techniques in predicting mean pressure coefficients and fluctuating pressure coefficients, but most of existing deep learning frameworks are limited to predicting a single snapshot using full spatial information. To forecast spatiotemporal wall pressure of flow past a rectangular cylinder, this study develops a physics-aware DeepU-Fourier neural Network (DeepUFNet) deep learning model. DeepUFNet comprises the UNet structure and the Fourier neural network, with physical high-frequency loss control embedded in the model training stage to optimize model performance. Wind tunnel testing was performed to collect wall pressures on two-dimensional rectangular cylinders using high-frequency pressure scanning, thereby constructing a database for DeepUFNet training and testing. The DeepUFNet model is found capable of forecasting spatiotemporal wall pressure information with high accuracy on the rectangular cylinder with side ratio 1.5. The comparison between forecast results and experimental data presents agreement in statistical information and physical interpretation. It is also found that embedding a physical high-frequency loss control coefficient $\beta$ in the DeepUFNet model can significantly improve model performance in forecasting spatiotemporal wall pressure information, particularly, high-order frequency fluctuation and wall pressure variance. Furthermore, the DeepUFNet extrapolation capability is tested with sparse spatial information input, and the model presents a satisfactory extrapolation ability. Last, the DeepUFNet is tested for generalization in unseen cases, rectangular cylinders with side ratio 4 and 3.75, and the model presents satisfactory generalization ability.

## I. INTRODUCTION

Wall pressure for rectangular cylinders is of great importance to understand the fluid-induced forces of rectangular cylinders,[1] structural responses,[2,3] civil structural safety, and building aerodynamics.[4,5] Wind tunnel testing[6–10] and numerical simulation[11–17] are considered as convincing approaches to obtain wall pressure information on rectangular cylinders. Existing research has accumulated a large amount of work that utilizes the wind tunnel testing method to collect and investigate wall pressure information for rectangular cylinders[18–20] as well as employing numerical tools to acquire and understand the rectangular cylinder wall pressure.[21–23]

As the new emerging technique spreads, artificial intelligence (AI) has achieved remarkable progress in temporal or spatiotemporal prediction of aerodynamic characteristics, showing advantages in the fitting of high-dimensional features[24–27] and the extraction of characteristics from the data.[28–32] For example, some researchers used AI techniques to predict the flow field or wind speed. Zhang *et al.*[33] introduced a hybrid deep learning model that combines a convolutional neural network (CNN)-long short-term memory network and microscale meteorological models to predict wind speed. Their approach explicitly integrates physical priors with data-driven components, improving both accuracy and interpretability in complex urban environments. Bastos *et al.*[34] harnessed the synergy between UNet and CNN to fuel spatiotemporal wind forecasting. In parallel, several studies explored graph-based and transformer-based architectures for spatiotemporal wind prediction. Pan *et al.*[35] proposed a spatiotemporal graph transformer network that effectively captured both local spatial and long-range temporal dependencies across multiple wind measurement stations.

As for pressure information prediction, some works employ artificial intelligence for wall pressure prediction and forecasting. Liu *et al.*[36] constructed two types of AI algorithms, graphic attention networks and dense neural networks, to make bidirectional predictions between the pressure of the rectangular cylinder wall and the wake flow. Their model evaluations on the numerical data showed that AI methods can predict both instantaneous wall pressure and time-averaged wall pressure based on the wake flow input. Tian *et al.*[37] developed a deep neural network (DNN) to predict mean and peak wall pressure coefficients on rectangular cylinder-shaped buildings. Based on the experimental data collected, the trained DNN model can make satisfactory predictions on the mean pressure coefficients. Fernández-Cabán *et al.*[38] built an artificial neural network (ANN) integrating backpropagation training to predict the mean, root mean square, and peak pressure coefficients on three geometrically scaled rectangular cylinders. Comprehensive experimental data were collected to train the model. Their evaluations presented that the ANN model was capable of predicting accurate mean, peak, and RMS pressure coefficients compared to the experimental data.

However, existing AI-based studies have focused mainly on using full-dimensional data to predict the instantaneous wall pressure distribution,[36] the mean pressure distribution,[39,40] the fluctuation pressure coefficient,[41] and the maximum pressure coefficient.[37] Few works have focused on spatiotemporal wall pressure forecast. In particular, there is an obvious absence of spatiotemporal wind tunnel wall pressure measurements forecast. This is partly due to the challenges posed by wind tunnel data, such as measurement sparsity, noise contamination, and non-stationary flow characteristics,[42–44] which make spatiotemporal and high-fidelity forecasting more difficult. Furthermore, in some practical scenarios and engineering cases, full-dimension wall pressure information is not always available or easily accessible, and there is a scarcity or noise in the measured data,[42–44] invalidating existing instantaneous wall pressure prediction models.

To address these limitations, we developed a hybrid deep learning framework, the physics-aware Deep U-Fourier neural network model (DeepUFNet), for robust spatiotemporal experimental wall pressure forecast. DeepUFNet is developed based on the UNet architecture and the Fourier neural network and embeds high-frequency loss control in the model training stage to optimize performance. As investigated in the literature, when the rectangular cylinder side ratio (SR) ranges from 1 to 2, the standard deviation value of the lift coefficient is very large, indicating a large fluctuation of wall pressure.[23] When $SR = 1.5$, the vortex generated from the leading corner is focused mainly on the regions around the upper and lower wall.[22] The visualized contours of instantaneous vortices show that the vortex near the upper and lower walls interacted with the vortex in the far wake, leading to complex flow characteristics near the upper and lower walls.[23,45,46] Therefore, during data collection, multiple rectangular cylinders are tested at different angle of attack in the wind tunnel to construct a large dataset for model training, and the side ratios range from 1.5 to 4 covering various flow patterns. After model training, we evaluated the model performance in spatiotemporal wall pressure forecast from different perspectives, including temporally averaged pressure information, power spectrum density of fluctuating pressure, instantaneous spatial distribution, and spatiotemporal correlation. In addition, the DeepUFNet model is tested with extrapolation capability with spatially sparse information input to model the data scarcity or corruption situations.

The general layout of this work is as follows. Section II presents the definition of the problem that is to be addressed in this study. Section III proposes the deep learning approach, the DeepUFNet architecture and details of the design. Section IV presents the experimental instruments, data acquisition, and data processing to generate the dataset for the deep learning model training and evaluation. The results and discussions are given in Sec. V. Furthermore, the DeepUFNet model extrapolation ability is tested with sparse spatial information input, which is shown in Sec. VI, and model generalization capability test results in shown in Sec. VII. Finally, the conclusion and future plans are expressed in Sec. VIII.

## II. PROBLEM SETUP

We define $\mathbf{\Omega} \subset \mathbb{R}^d$ as the spatial domain of the rectangular cylinder wall and $\mathbf{\Theta} \subset \mathbb{R}$ as the temporal domain. The wall pressure field is defined as a real-valued function: $\mathcal{P} : \mathbf{\Omega} \times \mathbf{\Theta} \to \mathbb{R}$, where $\mathcal{P}(\boldsymbol{x}, t)$ denotes the wall pressure at spatial location $\boldsymbol{x} \in \mathbf{\Omega}$ and time $t \in \mathbf{\Theta}$.

Given *a priori* pressure information $\mathcal{P}_{in}(\boldsymbol{x}, t)$ for $(\boldsymbol{x}, t) \in \mathbf{\Omega} \times \mathbf{\Theta}$, the objective of this work is to learn a high-dimensional nonlinear operator $\mathcal{G}_\theta$, such that

$$\mathcal{G}_\theta[\mathcal{P}_{in}(\mathbf{\Omega}, \mathbf{\Theta})] = \mathcal{P}_{out}(\mathbf{\Omega}, \mathbf{\Theta} + \Delta T), \tag{1}$$

where $\Delta T$ is the forecast temporal length, and it is 1000 snapshots in this work. $\mathcal{P}_{in}$ and $\mathcal{P}_{out}$ denote function spaces (e.g., $(\mathbf{\Omega} \times \mathbf{\Theta})$) of input and output spatiotemporal wall pressure fields, respectively. In this framework, $\mathcal{G}_\theta$ is parameterized by a deep learning model designed to approximate the mapping from input wall pressure distributions to future wall pressure distributions. The general goal of this work is to make spatiotemporal wall pressure forecast through a nonlinear operator $\mathcal{G}_\theta$ parameterized by a deep learning method.

## III. APPROACH: DEEPUFNET WITH PHYSICAL FREQUENCY EMBEDMENT

We develop a Deep U-Fourier neural Network (DeepUFNet) to reach the above spatiotemporal wall pressure forecast objective. The DeepUFNet model is developed based on the UNet framework and the Fourier neural network. UNet is a type of convolutional neural network originally designed for image segmentation and is one of the most classical deep learning techniques to generate feature projections from input to future information.[47–50] It's named UNet because of the U-shaped design of the network, which consists of the encoder part and decoder part. The encoder aims to downsample the data and increase the number of channels. The decoder is used to up-sample the data and decrease the number of feature channels. The method has been used for various engineering aspects and has shown extraordinary performance. Based on the original UNet framework, many re-developments have been performed to fit different practical engineering problems.[51–53] In the field of fluid dynamics, there are also many works utilizing the UNet model to predict fluid characteristics.[54–56] Fourier neural network (FNN) focuses on learning the mappings of the neural operator between function spaces,[57–60] such as the input projected to the output.[61,62]

In this work, UNet works as an encoder and decoder for the spatiotemporal wall pressure evolution. The FNN operates in the frequency domain, enabling separation of low- and high-frequency components. The constructed FNN is designed to model the high-frequency regime, which is particularly important for wind tunnel wall pressure measurements and often contains sensor noise and other high-frequency disturbances. By combining these two components, our approach aims to achieve stable, noise-resilient, and accurate multi-step spatiotemporal forecasts of wall pressure evolution. In addition, to make the model more robust and adaptive to noise, we embed different supplementary structures to improve the performance of the model. The detailed architecture design of the DeepUFNet model is shown in Fig. 1.

We define the DeepUFNet model as a composition of three main modules: a Fourier neural network for frequency-aware filtering and reconstruction, a UNet-based encoder-decoder for nonlinear spatiotemporal representation learning, and a linear projection layer for final projection to get the output. Formally, we express the overall model $\mathcal{G}_\theta$ as a composite operator,

$$\mathcal{G}_\theta = \mathcal{N}_o \circ \mathcal{N}_\Psi \circ \mathcal{N}_F, \tag{2}$$

where $\mathcal{N}_F : \mathcal{P}_{in} \mapsto \widetilde{\mathcal{P}}$ is the Fourier neural network, and it performs spectral decomposition, truncation, and neural reconstruction in the frequency domain,

$$\widetilde{\mathcal{P}} = \mathcal{F}^{-1}(\mathcal{N}_\Phi(\mathcal{F}(\mathcal{P}_{in}))), \tag{3}$$

where $\mathcal{N}_\Phi$ represents the network applied only to the high-frequency components. $\mathcal{N}_\Psi$ in Eq. (2) is the nonlinear operator based on UNet that performs the projection $\widetilde{\mathcal{P}} \mapsto \widetilde{\mathcal{P}}_o$. It captures hierarchical spatiotemporal features via convolutional encoders and decoders with skip connections. $\mathcal{N}_o$ in Eq. (2) is the output projection layer, which maps latent representations to forecast future pressure field $\widetilde{\mathcal{P}}_o \mapsto \mathcal{P}_{out}$. Consequently, the entire network performs the following mapping:

$$\mathcal{G}_\theta(\mathcal{P}_{in}) = \mathcal{P}_{out}. \tag{4}$$

Here, $\theta$ denotes all trainable hyperparameters across the model, and $\mathcal{P}_{out}$ is the forecast wall pressure field.

In detail, as illustrated in Fig. 1, the input to the model is denoted by $\mathcal{P}_{in}$, which represents the *a priori* temporal period of the wall pressure information in the function space $\mathbf{\Omega} \times \mathbf{\Theta}$. For implementation, this input is discretized as a matrix of shape $(m, n)$, where $m$ and $n$ correspond to the number of discretization points of temporal and spatial domains, respectively. In addition, the physical information of the rectangular cylinder, including the side ratio value (SR) and the wind angle of attack (AoA), also acts as input channels, resulting in three channels of input. The output has the same spatiotemporal shape $(m, n)$, but corresponds to a shifted time window in the temporal domain, i.e., $\mathbf{\Theta} + \Delta T$, with only one channel.

We then provide more detailed discussion of these three operators, $\mathcal{N}_o$, $\mathcal{N}_\Psi$, and $\mathcal{N}_F$ presented in Eq. (2). A Fourier neural network, shown in Fig. 1 blue part, denoted by $\mathcal{N}_F$, is applied to the input to approximate the mapping

$$\widetilde{\mathcal{P}} = \mathcal{N}_F(\mathcal{P}_{in}), \tag{5}$$

where $\widetilde{\mathcal{P}}$ denotes the FNN latent output in the DeepUFNet model, which will be transferred to UNet. The Fourier neural network begins by applying a Fast Fourier Transform (FFT) in the temporal domain of the input

$$f = \mathcal{F}(\mathcal{P}_{in}), \tag{6}$$

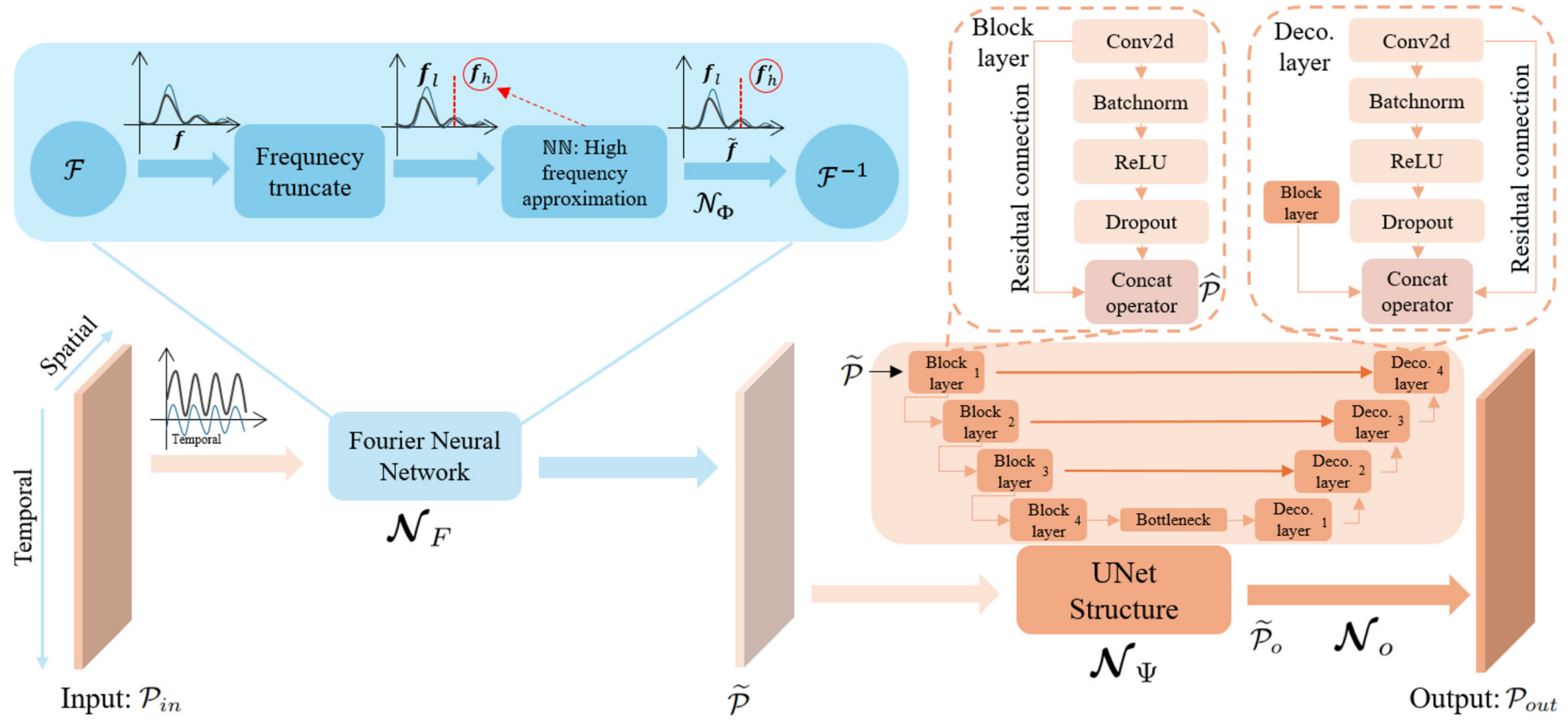

**FIG. 1.** DeepUFNet model architecture: the first part in blue is the Fourier neural network, and the second part in light orange is the UNet architecture. Within the UNet, there are block layers and decoder layers shown in the upper left zone. The input $\mathcal{P}_{in}$ and output $\mathcal{P}_{out}$ are all dimensionless values, where $\mathcal{P}_{in} = \mathcal{P}/(1/2\rho U^2)$.

where $\boldsymbol{f}$ denotes the temporal frequency representation of the input. Based on a predefined truncation frequency $f_t$, the frequency spectrum $\boldsymbol{f}$ is decomposed into a low-frequency part $\boldsymbol{f}_l$ and a high-frequency part $\boldsymbol{f}_h$. The low-frequency component $\boldsymbol{f}_l$ is retained unmodified (i.e., frozen), while the high-frequency component $\boldsymbol{f}_h$ is approximated by a neural network $\mathcal{N}_\Phi$ inside the FNN architecture as follows:

$$\widetilde{\boldsymbol{f}}_h = \mathcal{N}_\Phi(\boldsymbol{f}_h), \tag{7}$$

where $\mathcal{N}_\Phi$ is a multi-layer fully connected neural networks with trainable parameters, and $\Phi = [\phi_1, \phi_2, \phi_3]$, where $\phi_i$ denotes one layer of the fully connected networks. The network is initialized with random parameters drawn from a distribution with zero mean, and consists of three layers. Finally, the modified high-frequency component $\widetilde{\boldsymbol{f}}_h$ is combined with the unchanged low-frequency component $\boldsymbol{f}_l$, and the inverse FFT is applied to reconstruct the modified temporal signal in the physical domain.

Based on the Fourier neural network $\mathcal{N}_\Phi$ and approximated high-frequency component $\widetilde{\boldsymbol{f}}_h$, the entire frequency domain representation is updated from $\boldsymbol{f}$ to $\widetilde{\boldsymbol{f}}$.

Applying the inverse Fast Fourier Transform yields the updated pressure signal in the physical (temporal) domain,

$$\widetilde{\mathcal{P}} = \mathcal{F}^{-1}(\widetilde{\boldsymbol{f}}), \tag{8}$$

where $\widetilde{\mathcal{P}} \in \mathbb{R}^{m\times n}$ has the same shape as the input $\mathcal{P}_{in}$.

The output $\widetilde{\mathcal{P}}$ is then passed through a UNet architecture, shown in Fig. 1 light orange part, denoted by $\mathcal{N}_\Psi$, which consists of encoder-decoder structures with residual connections. Mathematically, this process is expressed as

$$\widetilde{\mathcal{P}}_o = \mathcal{N}_\Psi(\widetilde{\mathcal{P}}) = \prod_{j=1}^{n} \mathcal{N}_{\psi_j}(\widetilde{\mathcal{P}}), \tag{9}$$

where each $\mathcal{N}_{\psi_j}$ denotes a layer in the network, typically consisting of convolution, normalization, activation, and dropout. Here, $j$ ranges within n = 8 as shown in Fig. 1 light orange part, including four block layers and four decoding layers. The encoder contains four downsampling block layers shown in Fig. 1 left dashed line zone, and each block is composed of:

1. A 2D convolution with trainable weights and bias:

$$\boldsymbol{y}_1 = \mathcal{L}(\widetilde{\mathcal{P}}), \tag{10}$$

where $\mathcal{L}$ is a linear operator, that is a 2D convolution in this work.

2. A batch normalization layer:[63–65]

$$\boldsymbol{y}_2 = \frac{\boldsymbol{y}_1 - \mathbb{E}(\boldsymbol{y}_1)}{\sqrt{\mathrm{Var}(\boldsymbol{y}_1) + \epsilon}}, \tag{11}$$

where $\mathbb{E}(\cdot)$ and $\mathrm{Var}(\cdot)$, respectively, denote the mean and variance over local batch statistics, and $\epsilon$ is a small constant added for numerical stability.

3. A ReLU activation function:

$$\boldsymbol{y}_3 = \max(0, \boldsymbol{y}_2). \tag{12}$$

4. A dropout operation[66,67] with dropout rate $b = 0.3$:

$$\boldsymbol{y}_4 = \mathcal{D}(\boldsymbol{y}_3, b), \tag{13}$$

which randomly zeroes activations to improve generalization.

Residual connections[68–70] are incorporated through a concatenation operator $\mathcal{C}$,

$$\hat{\mathcal{P}} = \mathcal{C}(\widetilde{\mathcal{P}}, \boldsymbol{y}_4), \tag{14}$$

where the intermediate result $\hat{\mathcal{P}}$ is forwarded to the next layer or decoder stage. The decoder structure mirrors the encoder in symmetry, utilizing transposed convolution and concatenation with encoder features to enable feature fusion across scales. Finally, a linear projection without nonlinear activation maps the output to the forecasted pressure field in the future time interval,

$$\mathcal{P}_{out} = \mathcal{N}_o(\widetilde{\mathcal{P}}_o), \tag{15}$$

where $\mathcal{N}_o$ denotes the output layer of the DeepUFNet architecture, performing a final convolution.

For the entire DeepUFNet deep learning architecture design, the Fourier neural network is designed to capture the characteristics of high-frequency fluctuation through a high frequency approximation, and UNet is designed to make projection from input to output. For loss design, to better understand the capture of physical information of the machine learning technique, the loss of physical frequency is embedded during the model training stage. The mean squared error (MSE) loss is typically used to demonstrate the discrepancy between the forecast information and the ground truth. In this work, not only the MSE loss but also the high-frequency fluctuating feature discrepancy between the forecast result and ground truth is used to optimize the model and guide the model to understand the physics. In total, there are two parts of loss, including general MSE loss and physical frequency loss. The total loss is calculated as follows:

$$\mathcal{L} = \mathcal{L}_1 + \beta * \mathcal{L}_f, \tag{16}$$

where $\beta$ is a user-defined coefficient to adjust the influential factor of frequency loss, and $\beta$ is independent of time and frequency. To better guide the training of the DeepUFNet model, $\beta$ is set as a dynamic parameter to inform the model about the importance of physical frequency in different stages. $\beta$ is calculated as follows:

$$\beta = \beta_{set}\frac{epo}{n_{epo}}, \tag{17}$$

where $\beta_{set}$ is the user-set value, $epo$ is the current training epoch, and $n_{epo}$ are the total training epochs.

At the beginning of training, $\beta$ is very small, indicating that the model focuses primarily on minimizing the general MSE loss and capturing the dominant low-frequency features. As the epoch increases, $\beta$ grows linearly, gradually strengthening the contribution of the high-frequency loss term and enabling the model to learn finer-scale fluctuations. There are two folds of rationals for the $\beta$ setting. First, the linear scheduling has been widely adopted in deep learning to balance competing objectives in multi-term loss functions and progressively introduce more difficult tasks, which improves training stability and convergence.[71–73] Second, this progressive weighting aligns well with the scale hierarchy of bluff-body flows: large-scale coherent structures dominate the low-frequency content at the early stage, while small-scale vortical motions contribute to high-frequency pressure fluctuations at later stages.[74–77] To avoid excessive noise amplification, the truncation frequency is set at approximately three times the fundamental vortex-shedding frequency, where most of the energetic spectral content is concentrated. The initial value of $\beta$ is empirically chosen to match the order of magnitude of the baseline loss, ensuring balanced optimization between the low- and high-frequency terms. $\mathcal{L}_1$ is the MSE loss calculated from the ground truth and the forecast result as follows:

$$\mathcal{L}_1 = \frac{1}{k}\sum_{j=1}^{k}\left(\mathcal{P}_{fore,j} - \mathcal{P}_{gt,j}\right)^2, \tag{18}$$

where $k$ is the total dimension of temporal information, $\mathcal{P}_{fore,j}$ is the $j^{th}$ temporal pressure information of the forecast result $\mathcal{P}_{fore}$, and $\mathcal{P}_{gt,j}$ is the $j^{th}$ temporal pressure information of the ground truth $\mathcal{P}_{gt}$, that is experimental data in this work. $\mathcal{L}_f$ is the high frequency loss calculated from ground truth and forecast result as follows:

$$\mathcal{L}_f = \left\langle \left[\mathcal{F}_h(\mathcal{P}_{fore}) - \mathcal{F}_h(\mathcal{P}_{gt})\right]^2 \right\rangle, \tag{19}$$

where $\langle\rangle$ is the average in high frequency domain. To better capture high-frequency fluctuation, the frequency loss $\mathcal{L}_f$ is designed to optimize the direction of model training development. Similar to the Fourier neural network, there is a frequency truncation $f_t'$ in the training of the model for the calculation of $\mathcal{L}_f$. After performing a fast Fourier transform $\mathcal{F}$ on both $\mathcal{P}_{fore}$ and $\mathcal{P}_{gt}$, the frequency domains results are collected as $\mathcal{F}(\mathcal{P}_{fore})$ and $\mathcal{F}(\mathcal{P}_{gt})$. We then truncate $\mathcal{F}(\mathcal{P}_{fore})$ and $\mathcal{F}(\mathcal{P}_{gt})$ by preserving results with frequency higher than the frequency truncation $f_t'$ (i.e., $f > f_t'$), which are, respectively, denoted as $\mathcal{F}_h(\mathcal{P}_{fore})$ and $\mathcal{F}_h(\mathcal{P}_{gt})$. When the truncation frequency was determined, we investigate the existing literature for the rectangular cylinder studies[78] and our flow field observations. In bluff-body wakes, the dominant energy concentrates at the fundamental dominant vortex-shedding frequency $f_s$[79,80] and its low-order harmonics (around 2-3$f_s$),[81] whereas higher-frequency energy decays rapidly.[82] Therefore, we set the truncation at 110 Hz, which is slightly larger than $3f_s$ to prioritize dynamically significant content while avoiding noise amplification and unstable gradients at high frequencies.

Based on the construction of the model and the preparation of data, the model is trained on the data collected from the wind tunnel experiment. In model training, the hyperparameters of the DeepUFNet architecture and truncation frequencies $f_t$ and $f_t'$ are set as shown in Table I. In the model training stage, to avoid the phenomenon of overfitting and simultaneously enhance the generalization ability of the model, several actions are taken, including introducing weight decay[83,84] to adjust the learning rate of the model in the training, using the dropout technique[66,67] shown in Eq. (13) to lose some random information in neural networks, and applying batch normalization shown as Eq. (11) (Refs. 63–65) in the DeepUFNet model.

**TABLE I.** Hyperparameters in the model and training stage.

| Hyperparameters | Values |
|---|---|
| Batch size | 16 |
| $n_{epo}$ | 2000 |
| Learning rate | 0.0001 |
| Weight decay rate | 0.0001 |
| Dropout rate | 0.3 |
| $\beta_{set}$ | 0.001 |
| $f_t$ in $\mathcal{N}_F$ and $f_t'$ in $\mathcal{L}_f$ | 110 Hz ($S_t = 0.33$) |
| Loss setting | MSE loss and physical frequency loss ($\beta$ embedding) shown in Eq. (16) |

## IV. DATA GENERATION AND PROCESSING

### A. Experimental setups and instruments

The experiment was carried out in the Artificial Intelligence for Wind Engineering (AIWE) Laboratory of the Harbin Institute of Technology, Shenzhen. The wind tunnel is a closed-loop wind tunnel with a test section of 800 × 500 × 500 mm$^3$ (length × width × height) shown in Fig. 2(a). In this experiment, the approaching flow velocity is set at 10 m/s, and no turbulent grid is set in front of the test section. The turbulence intensity of the wind tunnel is less than 0.4% evaluated with the *Hanghua CTA04-EDU* hot-wire, indicating that the approaching flow is treated as uniform flow.[20,85] At the beginning of each experiment, to ensure the approaching flow to the rectangular cylinder strictly follows zero angle of attack, we calibrated the rectangular cylinder arrangement with a 12-line laser calibration instrument before the experiment, as shown in Fig. 2(b). In this experiment, the pressure scanning starts working after the approaching flow is consistent and the turbulent flow has been fully developed in the rectangular cylinder wake. Multiple rectangular cylinders are tested in the wind tunnel under various wind angle of attacks (AoAs). The rectangular cylinder and dimensional information are presented in Table II. Due to the effect of the wind tunnel blockage ratio, the tested AoAs vary for different rectangular cylinders. The Reynolds numbers ($Re$) of these cases range from $1.35 \times 10^4$ to $4.06 \times 10^4$.

### B. Experimental data validation

The pressure scanners used in the experiment to sample wall pressure information are *Scanivalve MPS4262* pressure scanners. In this experiment, the experimental environment condition is around 26 °C, and the experimental temperature remains the best working condition for the pressure scanning system,[40] which is monitored and controlled by a constant air conditioner. The length of the pressure tube is less than 0.6 m to reduce the distortion of wall pressure information in the experiment. High-frequency pressure scanning with sampling frequency $f_s = 400$ Hz is used to sample wall pressure.

Figure 3 presents the distribution of the pressure valves on the wall and the dimension information for the rectangular cylinder of side ratio 1.5. On both upper and lower walls, there are eight pressure taps. There are five pressure taps on the left and right, that is, the leading and trailing walls. For all rectangular cylinders, on four walls, the distance between each pressure tap is 5 mm. The pressure sampled by the pressure scanners is relative pressure, indicating that the pressure obtained has been subtracted by the reference pressure $p_{ref}$. For each experiment, the total number of temporal snapshots collected for wall pressure is 24 000. More information on the wind tunnel and experimental devices can be found in the work of the AIWE lab.[86,87]

With the collected pressure information on the wall of the rectangular cylinder, the pressure information is processed to validate the

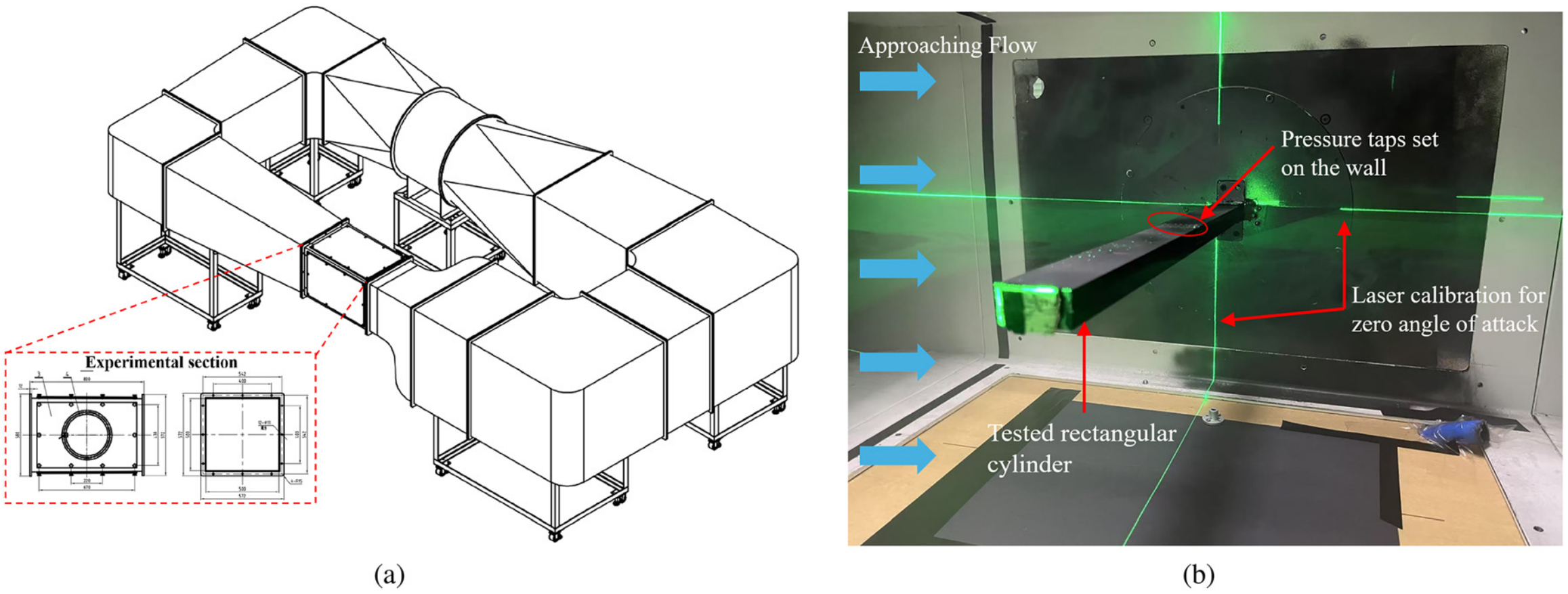

FIG. 2. Experimental instruments and rectangular cylinder dimensional information: (a) Closed-loop wind tunnel in AIWE Laboratory, (b) wind tunnel testing physical settings and calibrations. Green lines are the zero angle of attack calibration using a 12-line laser calibration instrument.

TABLE II. Pressure taps arrangement and dimension information for all rectangular cylinders.

| Side ratio | $L$ (mm) | $D$ (mm) | No. of pressure taps on the length | No. of pressure taps on the width | AoA |
|---|---|---|---|---|---|
| 1.5 | 45 | 30 | 8 | 5 | 0°–90° with 5° interval |
| 1.75 | 35 | 20 | 6 | 3 | 0°–90° with 5° interval |
| 2 | 40 | 20 | 7 | 3 | 0°–90° with 5° interval |
| 2.25 | 45 | 20 | 8 | 3 | 0°–90° with 5° interval |
| 2.5 | 50 | 20 | 9 | 3 | 0°–90° with 5° interval |
| 2.75 | 55 | 20 | 10 | 3 | 0°–90° with 5° interval |
| 3 | 60 | 20 | 11 | 3 | 0°–90° with 5° interval |
| 3.25 | 65 | 20 | 12 | 3 | 0°–45° with 5° interval |
| 3.5 | 70 | 20 | 13 | 3 | 0°–40° with 5° interval |
| 3.75 | 75 | 20 | 14 | 3 | 0°–35° with 5° interval |
| 4 | 80 | 20 | 14 | 3 | 0°–30° with 5° interval |

precision of the experimental data by comparing the collected information with existing research data. For flow past a rectangular cylinder under the angle of attack 0°, the Strouhal number ($St$), the mean drag coefficient ($\overline{C}_d$), and the standard deviation value of the lift coefficient

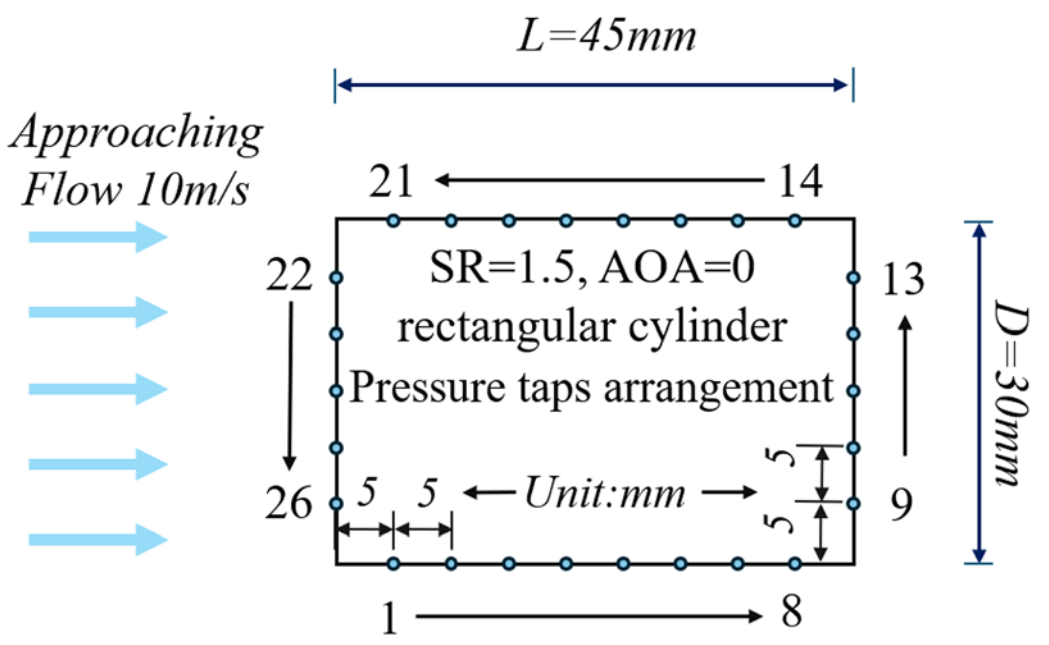

FIG. 3. Rectangular cylinder of side ratio 1.5 dimension information and pressure taps arrangement. The number outside the rectangular cylinder is the pressure tap index, and the number inside the rectangular cylinder denotes the spatial distance (5 mm) between each pressure tap. The pressure taps are uniformly distributed.

($C_{l,std}$) determine the precision of the pressure information. Thus, these aerodynamic statistical features are calculated to validate the experimental data collected from the wind tunnel test. The mean drag coefficient $\overline{C}_d$ is calculated as follows:

$$\overline{C}_d = \frac{1}{N}\sum_{i=1}^{N} C_{d,i}, \tag{20}$$

where $N = 24,000$ is the total number of snapshots collected from the wind tunnel experiment. $C_{d,i}$ is the instantaneous drag coefficient, and the instantaneous drag coefficient is calculated by the instantaneous wall pressure difference on both leading and trailing walls of the rectangular cylinder as follows:

$$C_{d,i} = \frac{1}{n_d}\sum_{j=1}^{n_d} \frac{p^i_{j,r} - p^i_{j,l}}{1/2\rho U^2}, \tag{21}$$

where $n_d = 5$ is the number of pressure taps on the leading and trailing walls for the rectangular cylinder of the side ratio 1.5 in this study. $U = 10\,\text{m/s}$ is the approaching wind velocity in this experiment. Here, $\rho = 1.225\ \text{kg/m}^3$ is the air density. $p^i_{j,l}$ and $p^i_{j,r}$, respectively, denote the wall pressure information sampled by the $j$th spatial pressure scanners on leading and trailing walls at the particular temporal instant $i$. Here, $i$ ranges in 24 000, which is the entire temporal domain. It should be noted that the instantaneous darg coefficient ($C_{d,i}$) here is the pressure difference of the leading and trailing walls, without considering the friction drag under this condition, as when the Reynolds number is greater than 200, the friction drag can be neglected for rectangular cylinder at $SR = 1.5$.[14] The instantaneous lift coefficient is calculated as follows:

$$C_{l,i} = \frac{1}{n_l}\sum_{j=1}^{n_l}\frac{p^i_{j,low} - p^i_{i,up}}{1/2\rho U^2}, \quad (22)$$

where $n_l = 8$ is the number of pressure taps on the upper and lower walls, $p^i_{j,low}$ and $p^i_{i,up}$ denote the wall pressure information sampled by the pressure tap at the $j$th spatial pressure tap on the upper and lower wall at the particular time instant $i$, and $i$ ranges within 24 000. After getting the instantaneous lift coefficient $C_{l,i}$, the standard deviation of the lift coefficient can be calculated based on the temporal domain as follows:

$$C_{l,std} = \sqrt{\frac{1}{N}\sum_{i=1}^{N}(C_{l,i} - \overline{C}_l)^2}, \quad (23)$$

where $N = 24,000$ is the total number of snapshots as in Eq. (20), and $\overline{C}_l$ is the temporal average of the lift coefficient $C_{l,i}$. Moreover, the fast Fourier transform (FFT) is performed based on the temporal history of lift coefficient to obtain the wake vortex shedding dominant frequency $f$, which is used to calculate the Strouhal number ($St$). The Strouhal number is calculated as follows:

$$St = \frac{fD}{U}, \quad (24)$$

where $D = 0.03$ m is the characteristic length for the rectangular cylinder of side ratio 1.5 and $U = 10$ m/s is the approaching velocity. Based on the wall pressure information collected, the dominant vortex shedding frequency is calculated as around 34.3 Hz. The pressure scanning frequency of 400 Hz can be sufficient to capture fluctuating features. Moreover, the Reynolds number is defined as

$$Re = \frac{UD}{\nu}, \quad (25)$$

where $\nu = 1.51 \times 10^{-5}$ m$^2$/s is the kinematic viscosity of the air. The above three aerodynamic characteristics, including $St$, $\overline{C}_d$, and $C_{l,std}$, can be compared with historical research data, as shown in Table III.

**TABLE III.** Aerodynamic statistics ($St$, $\overline{C}_d$, $C_{l,std}$) comparison between wind tunnel experimental data here and existing research data for the rectangular cylinder of side ratio 1.5 under the angle of attack 0°.

| Literature | $Re$ | $St$ | $\overline{C}_d$ | $C_{l,std}$ |
|---|---|---|---|---|
| Shimada and Ishihara[22] | 24 000 | 0.103 | 1.62 | 0.50 |
| Sohankar[23] | 100 000 | 0.095 | 1.62 | ··· |
| Wang *et al.*[88] | 685 000 | ··· | 1.63 | 0.67 |
| This work | around 20 000 | 0.103 | 1.64 | 0.51 |

When comparing aerodynamic statistics with those in existing literature, it can be seen that the wall pressure information collected from this wind tunnel experiment is in good agreement with existing research data[22,23,88] for the rectangular cylinder of side ratio 1.5 in this study. Statistical verification indicates that the wall pressure measurements can serve as a basis for further DeepUFNet model training and testing in Sec. V.

### C. Data preparation for model training

After validating the high-frequency pressure scanning information collected from the wind tunnel experiment, the pressure data are processed to feed to the deep learning model. As mentioned, the objective of this work is to achieve the spatiotemporal wall pressure forecast goal. Thus, during data preparation for the deep learning model, wall pressure information in the temporal domain should strictly follow the sequence. Data sampling and data division for DeepUFNet training and testing are shown in Fig. 4. The entire database is divided into the training dataset and the testing dataset. The training dataset comprises 70% of the total database, and the last 30% forms the testing dataset. As presented in the experimental model section, there are 26 pressure taps on the walls of the rectangular cylinder. The total spatiotemporal data forms a matrix of size (24 000, 26). The entire dataset is normalized in the spatiotemporal domain, fitting the data in the range (0, 1), and the data normalization aims to assist the DeepUFNet model convergence in the later model training stage. Data normalization is performed based on the following equation:

$$C_{p,norm} = \frac{C_p - C_{p,min}}{C_{p,max} - C_{p,min}}, \quad (26)$$

where $C_{p,min}$ and $C_{p,max}$ are the minimum and maximum wall pressure coefficients in the entire wall pressure coefficient dataset of size (24 000, 26), respectively. $C_p$ is the instantaneous temporal wall pressure coefficient at a particular spatial point, which is calculated as follows:

$$C_p = \frac{p - p_{ref}}{1/2\rho U^2}, \quad (27)$$

where $p$ is the sampled pressure information with unit $pa$, $p_{ref}$ is the reference pressure, which is the same throughout the experiment with unit $Pa$, $\rho$ and $U$ are the air density and the approaching velocity, respectively, which are the same as in Eq. (21). Note that in this experiment, the reference pressure is subtracted in the calibration stage. Therefore, the pressure information collected is the target wall pressure. As presented in the above subsection of Table II, for rectangular cylinders with different side ratios, the number of pressure taps on the wall varies due to the difference in dimensions. Therefore, we make the dataset size consistent before model training. For all rectangular cylinders, we interpolate or reduce the number of pressure taps on the length as 8, and interpolate the number of pressure taps on width as 5, leading to in total 26 pressure taps on four walls.

As shown in Fig. 4, $p^i_j$ denotes the wall pressure information at the spatial $j$th pressure tap for the temporal $i$th snapshot. Here, $i$ ranges within 24 000, and $j$ ranges within 26. The entire dataset is divided into training and testing subsets, and the first 70% of temporal wall pressure snapshots with full 26 spatial points measurements are used to train the model, and the other 30% of temporal snapshots with full 26 spatial points measurements are used to test the model performance. In this work, the temporal length of the forecast is set as 1000 snapshots and the spatial information of the forecast is the full spatial measurements, indicating that the input size and forecast are identical to (1000, 26). The model training hardware is NVIDIA RTX 4090Ti GPU. The training time for the DeepUFNet model is around 10 h.

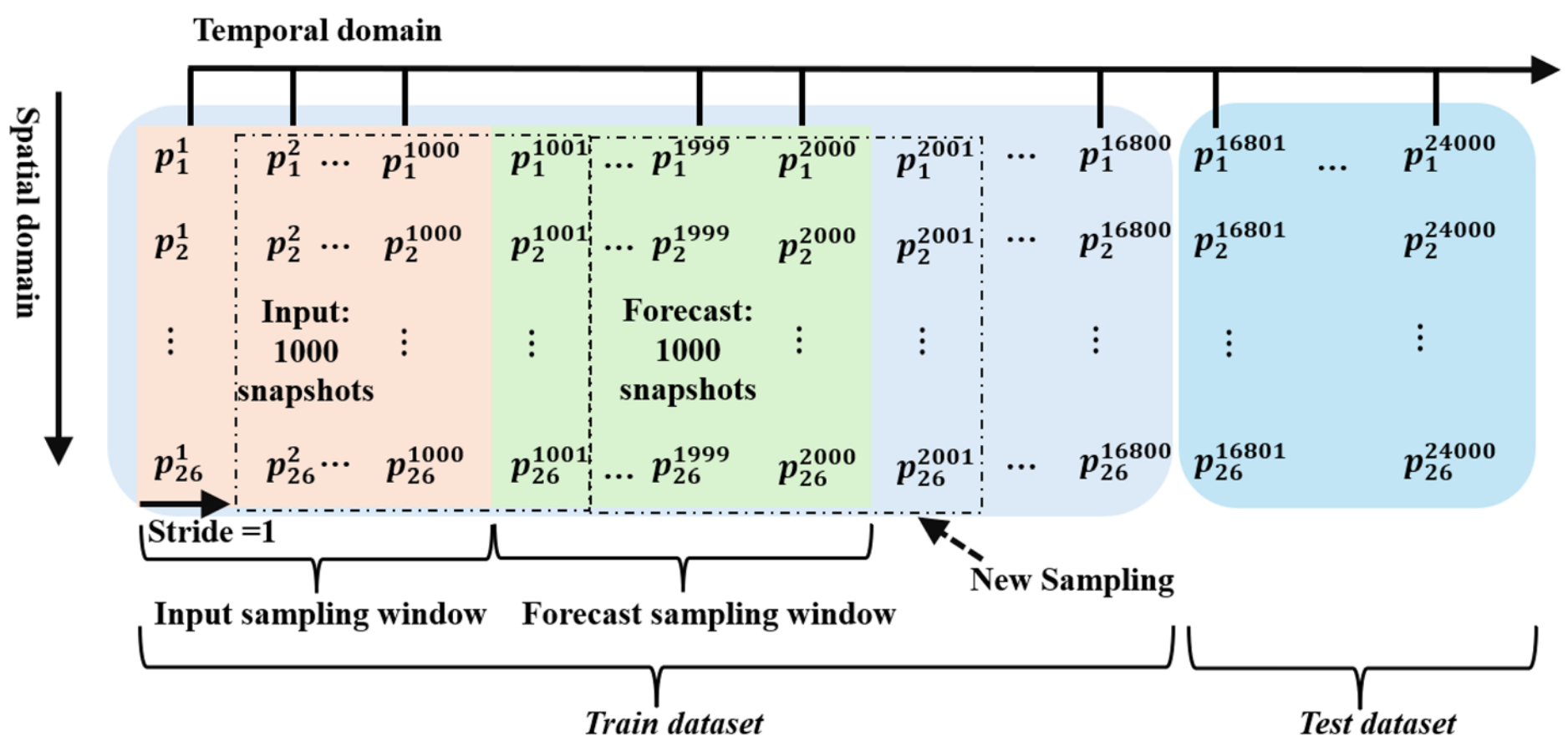

FIG. 4. Wall pressure database division training and testing for deep learning and data sampling flow chart to get input information and the corresponding forecast wall pressure information: 70% of the entire dataset is used for training, and the other 30% is used for testing. The data sampling stride is 1, and the temporal length of both input and forecast is 1000.

## V. FORECAST PRESSURE RESULTS AND ANALYSIS

After model training, the DeepUFNet model is applied to forecast spatiotemporal wall pressure information for the rectangular cylinder. As mentioned above, the target forecast timescale length is 1000 snapshots at 26 spatial points on the wall. In this section, we demonstrate the model performance on forecast wall pressure for the rectangular cylinder of side ratio 1.5, and Sec. VII will present the model generalization capability in forecast wall pressure for other cases. Indicated by the data preparation procedures shown in Fig. 4, there are a total of 5200 pairs of data in the test dataset. We evaluate the model performance and mean squared error (MSE) loss distribution over the 5200 pairs of data in the test dataset. Figure 5 presents the MSE loss distribution on the entire test dataset. From the MSE distribution shown in Fig. 5, we selected the data pair No. 50 with $MSE = 0.0302$ as the input wall pressure to the DeepUFNet model, which shows the highest number of occurrences.

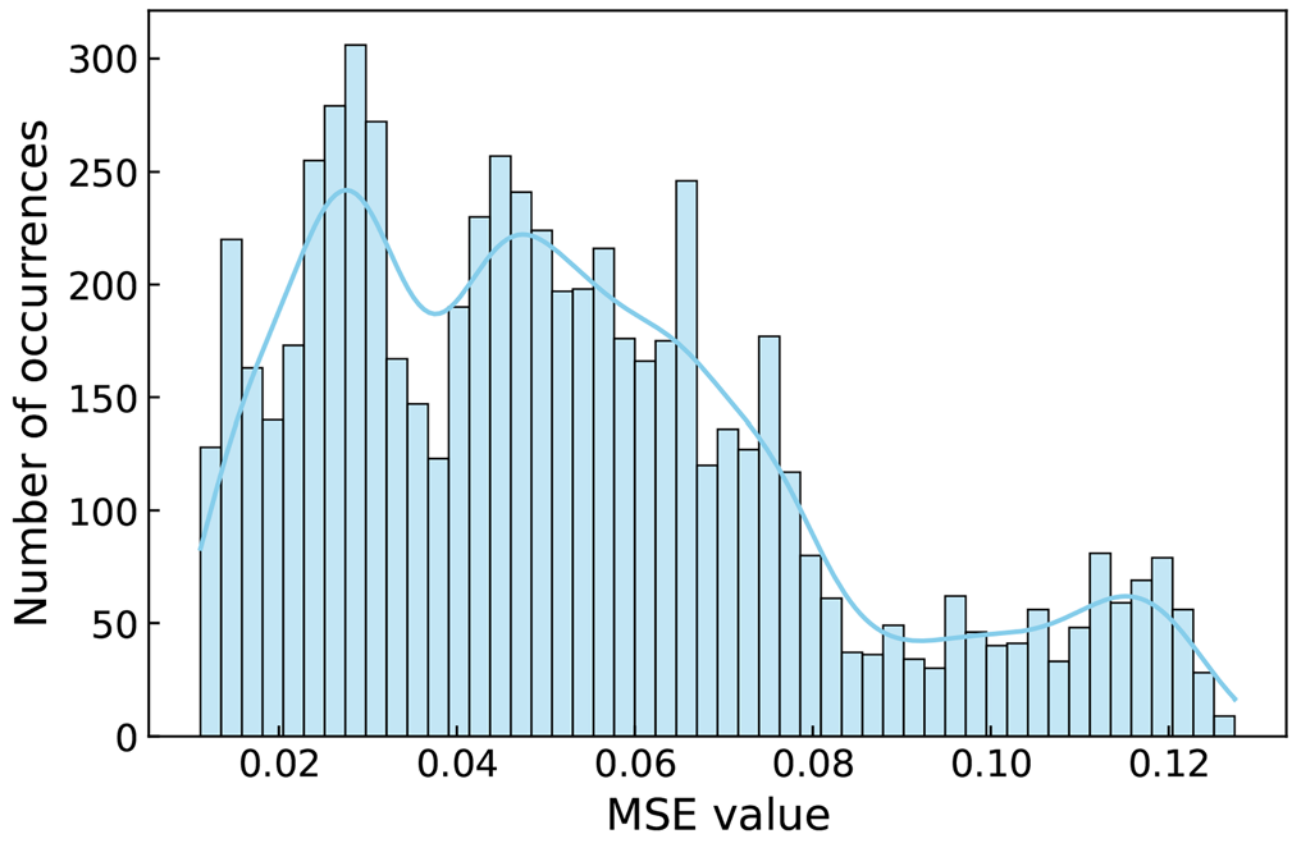

FIG. 5. Mean squared error (MSE) distribution of model performance on the entire test dataset. x axis is the MSE value, and y axis is the number of occurrences.

In other words, the physical time snapshots of the input are snapshot Nos. 16 850–17 849, and the target forecast information are wall pressure of snapshot Nos. 17 850–18 849 as indicated in Fig. 4. In the following parts of model performance estimation, the demonstration results are evaluated on the basis of this selected data pair.

### A. Baseline comparison

Based on the above model and parameters settings, we compare the DeepUFNet performance and baseline models performance, including UNet and FNO model. For the UNet model, we use eight layers of convolutional layers, which are presented in right hand side of Fig. 1. For the FNO model, we use identical architecture as shown in left hand side of Fig. 1. Similarly, we use NVIDIA RTX 4090Ti GPU to train the model. As stated above, the number of 50 pair of data in the dataset is set as input to evaluate the model performance for the same prediction target. The forecast wall pressure is the subsequent 1000 snapshots of wall pressure information. We train the baselines with identical loss settings shown in Eq. (16) and quantify the MSE loss on this particular input based on Eq. (18). Table IV presents the MSE values of three different models. From the baseline comparison, it can be seen that the DeepUFNet model has lower MSE value on the spatiotemporal wall pressure forecast, indicating its advantages over other two baselines. In the following results, we only show prediction from DeepUFNet model.

### B. Statistical results

The temporally averaged statistical wall pressure information is first investigated at these 26 spatial points, that is, $\overline{C}_p$ on the wall. After collecting the forecast wall pressure, the temporally averaged wall pressure coefficient $\overline{C}_p$ is calculated as below:

$$\overline{C}_{p,j} = \frac{1}{n}\sum_{i=1}^{n} C^{i}_{p,j}, \tag{28}$$

where $\overline{C}_{p,j}$ denotes the temporally averaged pressure coefficient at the $j$th spatial pressure tap, and $C^{i}_{p,j}$ denotes the $i$th instantaneous temporal pressure coefficient at $j$th spatial pressure tap. In this case, $i$ ranges within $n = 1000$, and $j$ ranges within 26. In addition, the standard deviation of the wall pressure information ($\sigma$) is investigated at every spatial pressure tap to study the temporal fluctuations.

**TABLE IV.** MSE comparison based on the No. 50 data pair input for different models.

| Model Type | MSE value |
|---|---|
| UNet | 0.0373 |
| FNO | 0.0529 |
| DeepUFNet | 0.0302 |

As shown in Fig. 6, the vertical bar denotes the standard deviation ($\sigma$) for each spatial pressure tap in both the forecast and experimental data. The circular point denotes the temporally averaged wall pressure coefficient for the forecast result with high-frequency loss control coefficient $\beta$, and the square point denotes the temporally averaged wall pressure coefficient information for the experimental data. In addition to the two sets of data above, another forecast information is presented without embedding $\beta$ in the model training stage (i.e., $\beta = 0$) to compare model performance. This comparison is aimed at demonstrating the positive effects of the physical high-frequency loss control coefficient $\beta$ in enhancing model performance. The $y$ axis denotes the wall pressure fluctuation range as $\overline{C}_p \pm \sigma$.

From the results shown in Fig. 6, it can be seen that for the temporally averaged wall pressure coefficient $\overline{C}_p$, both the forecast results, that is, with or without $\beta$ embedment in the model training stage, align with the experimental data, indicating that the DeepUFNet model can forecast the temporally averaged wall pressure coefficient information. For most pressure taps on the rectangular cylinder walls, the standard deviations of the experimental data ($\sigma_m$) are slightly larger than the standard deviation values of two kinds of forecast results. In addition, it can be seen that the standard deviation value ($\sigma_1$) of the forecast result (with $\beta$) is higher than the standard deviation value ($\sigma_2$) of the forecast result (without $\beta$). The above information indicates that the DeepUFNet model with embedding a dynamic coefficient $\beta$ in the physical frequency loss part can improve the performance in forecasting temporal variation and fluctuation of the wall pressure information. In particular, on both the leading and the trailing wall, that is, pressure tap Nos. 22–26, and 9–13, the standard deviations of both the forecast results $\sigma_1$ (with $\beta$) and $\sigma_2$ (without $\beta$) are almost identical to the standard deviation value of the experimental data $\sigma_m$. This phenomenon can be understood as the wall pressure on the leading wall has less fluctuation in temporal domain, and the wall pressure on the leading wall is mainly induced by the approaching flow. In the trailing wall, the wall pressure is highly correlated with the periodic vortex shedding, and periodic information can be excellently captured by the deep learning model from the frequency domain.[89–91] On the upper and lower walls, that is, pressure tap Nos. 1–8, 14–21, $\sigma_1$ and $\sigma_2$ are slightly lower than $\sigma_m$. This phenomenon can be explained by the flow field information around the upper and lower walls of the rectangular cylinder being very complex, and the flow complexity leads to a highly fluctuating wall pressure.[16,22] From statistical information shown in Fig. 6, it can be derived that the DeepUFNet model can forecast temporally averaged wall pressure information, and embedding a dynamic coefficient $\beta$ in the DeepUFNet model improves the model's ability to forecast temporal pressure variation. In the later part of this work, if there is no additional note on the forecast wall pressure, the forecast result denotes the forecast results using DeepUFNet model trained with embedding $\beta$.

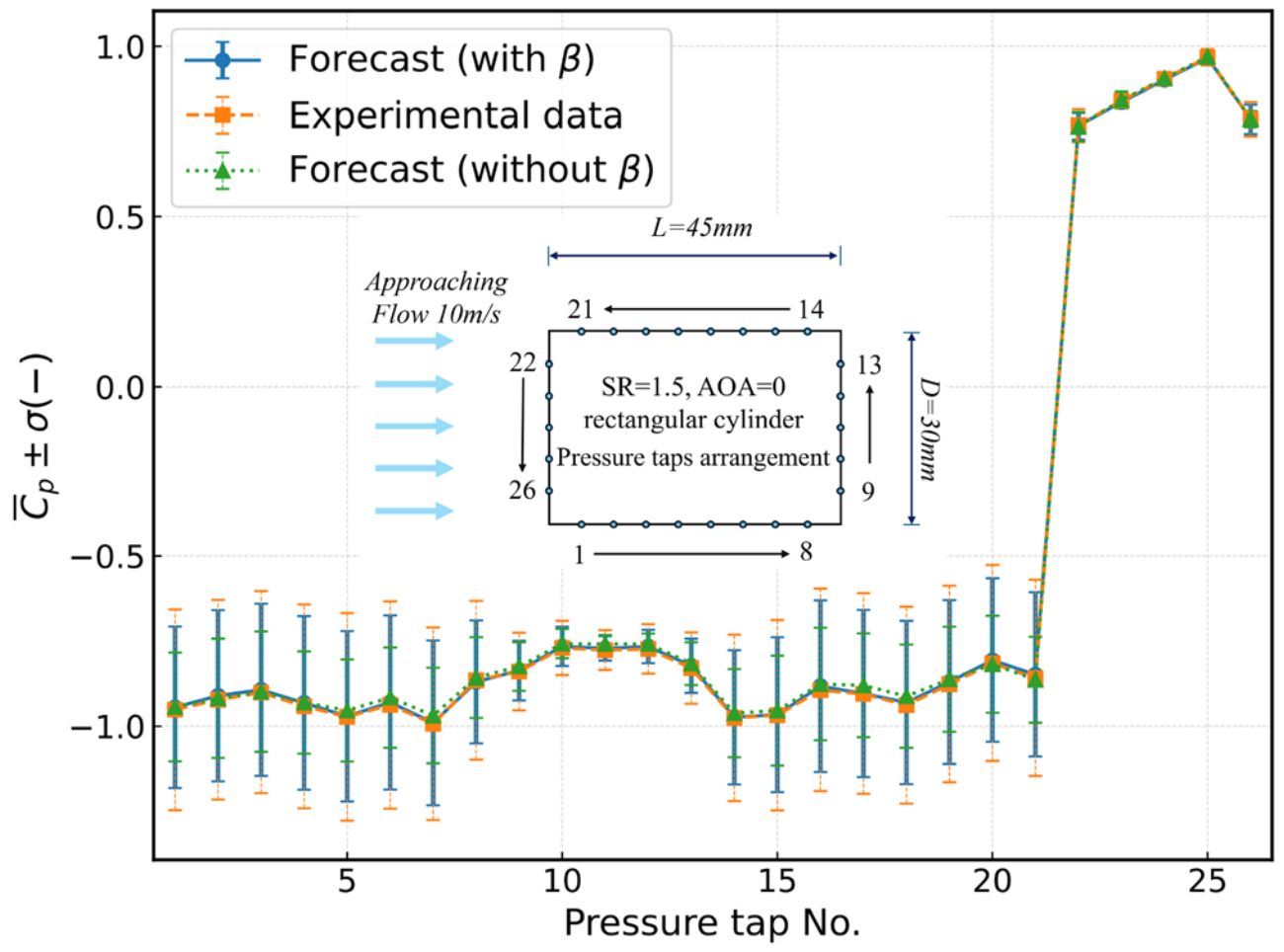

**FIG. 6.** Statistical information comparison between the forecast result and experimental data including the temporally averaged wall pressure coefficient $\overline{C}_p$ in different spatial points, and the standard deviation value $\sigma$ of temporal wall pressure information at different spatial points. Scatter points denote temporally averaged wall pressure coefficient $\overline{C}_p$, and vertical bars denote the standard deviation $\sigma$. In y axis, the sign (-) denotes the dimensionless values of $\overline{C}_p \pm \sigma$.

Based on statistical information and the standard deviation difference shown in Fig. 6, four points on the rectangular cylinder walls are selected to demonstrate the performance of the model: point 5 around the middle part of the lower wall, point 9 in the lower part of the trailing wall, point 18 on the upper wall, and point 22 on the upper part of the leading wall. Figure 7 shows the forecast information (with $\beta$) and the experimental data from the temporal time series at these selected pressure taps. The temporal time-series information presented in Fig. 7 shows a slight discrepancy between the forecast results and the experimental data. Overall, the DeepUFNet model can capture the characteristics of oscillation. However, in some peaks, the DeepUFNet model cannot perfectly fit the peak pressure information. Regarding the x axis of Fig. 7, $\Delta T$ denotes the physical time information accumulating from the start of the forecast with unit *s*. $\delta t$ denotes the pressure scanning interval, which is $1/f_s$ in the experiment with unit *s*. Therefore, the x axis $\Delta T/\delta t$ shown in Fig. 7 is a dimensionless parameter, indicating the snapshot number.

To estimate the performance of the model in forecasting the spatiotemporal wall pressure information, the statistical information of the forecast wall pressure in the temporal domain is studied and the probability density function (PDF) is drawn. We select four pressure tap Nos. 5, 9, 18, and 22, which are the same spatial locations in Fig. 7 selected to demonstrate the DeepUFNet performance. The probability density function is calculated on the temporal pressure information at these particular spatial points.

As shown by the PDF presented in Fig. 8, the forecast results (with $\beta$) have peaks and valleys similar to the experimental data, while the magnitudes of the peaks and valleys are different. Presented in Fig. 8, it can be seen that the forecast results have a higher probability concentration on the pressure information around the peaks. For example, for pressure tap 9 located in the lower part of the trailing wall, there is a mountain peak around $C_p = -0.78$ as shown in Fig. 8(b). The DeepUFNet model can capture the peak at this specific $C_p$ value, and the forecast result has a PDF value higher than that of the experimental data. In addition, in zone $C_p \in (-1.0, -0.9)$ of Fig. 8(b), the model can capture the changes of the PDF gradient. For pressure tap Nos. 5, 18, and 22, shown in Figs. 8(a), 8(c), and 8(d), there are two peaks in the PDF distribution for both the forecast result and the experimental data, and the forecast result has peak values at $C_p$ close to the experimental data. For example, for pressure tap 5, there are two peaks at $C_p \approx -1.3$ and $-0.6$ for the experimental data. The forecast wall pressure has two similar peaks.

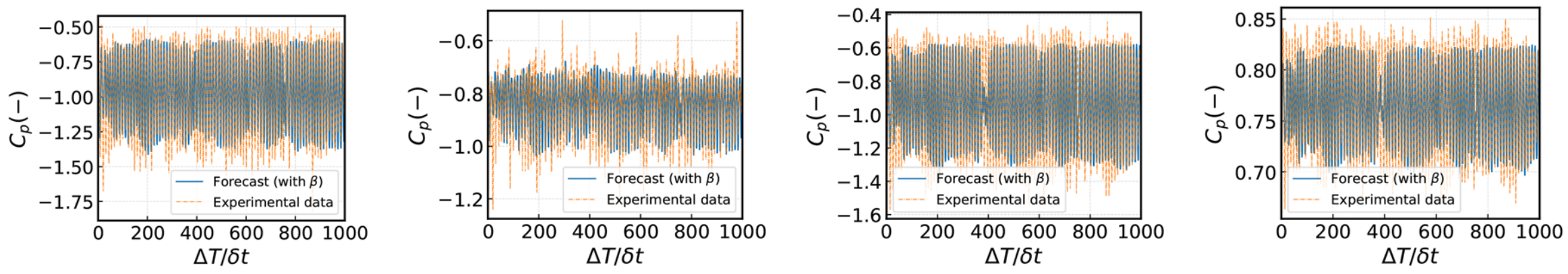

(a) (b) (c) (d)

**FIG. 7.** Temporal wall pressure information comparison between forecast result (with $\beta$) and experimental data at four particular spatial points, (a) pressure tap 5 on the lower wall, (b) pressure tap 9 on the trailing wall, (c) pressure tap 18 on the upper wall, and (d) pressure tap 22 on the leading wall. The sign (-) in y axis denotes the dimensionless values of the pressure coefficient. In x axis, $\Delta T$ denotes the physical time information accumulating from the start of forecast with unit *s*, $\delta t$ denotes the pressure scanning time interval with unit *s*.

### C. Fluctuating pressure power spectrum density

In this part, the power spectrum density (PSD) of the temporal fluctuating pressure information is used to compare the fluctuation characteristics on two sets of the forecast result (with $\beta$ and without $\beta$) with experimental data. The PSD of fluctuating pressure information is calculated based on Fast Fourier Transform. In order to obtain nondimensional parameters to demonstrate the pressure features in the temporal domain, the frequency $f$ is nondimensionalized to the Strouhal number with Eq. (24). Similar to Sec. V B, the results of the fluctuating pressure power spectrum density at four spatial points, including pressure tap Nos. 5, 9, 18, and 22, are calculated and shown in Fig. 9. As demonstrated in Table I, the truncation frequency $f_t'$ is 110 Hz, which corresponds to $St = 0.33$ shown in the figure with a "frequency cutoff."

From the results shown in Fig. 9, the distribution of the power spectrum density of the forecast results (with $\beta$) at these four spatial points is much closer to the power spectrum density distribution of the experimental data than the forecast results without $\beta$. In particular, comparing two types of forecast results with $\beta$ and without $\beta$, it can be found that when embedding the value $\beta$ in model training, the peak distribution of the power spectrum density can be excellently captured, including the dominant frequency $St_m$, twice of the dominant frequency $2St_m$, and three times of the dominant frequency $3St_m$ shown in all spatial pressure taps of Fig. 9. Specifically, for the pressure tap No. 9, which is located in the trailing wall, as shown in Fig. 9(b), three peaks are detected by the DeepUFNet model with embedding $\beta$, and three peaks are very close to the experimental data. The above findings indicate that DeepUFNet can forecast both low-frequency regime and high-frequency regimes. Although the high-frequency loss $\mathcal{L}_f$ calculated from Eq. (19) is imposed for frequency higher than $f_t'$ approximately three times of the dominant frequency, the overall model forecast performance is improved for both the low-frequency and high-frequency regimes. That is, adding a physical high-frequency loss term $\mathcal{L}_f$ associated with $\beta$ in the total loss control in Eq. (16) can greatly enhance the model performance. Nevertheless, from the presented comparison between forecast result and experimental data shown in Figs. 9(c) and 9(d), the peak bandwidths of the forecast result are wider than those of experimental data. The experimental data have sharp peaks with narrow bandwidths, particularly near the dominant frequency $St_m$, indicating some minor discrepancies. Furthermore, in the high-frequency regime, two types of forecast results present a difference compared to the experimental ground truth, and the forecast result (with $\beta$) is closer to the ground truth. This phenomenon can be explained from the perspective of fluid flow: the low-frequency regime is associated with large vortex structures, which can be well captured, while the high-frequency regime is associated with small dissipation vortex, which are difficult to capture.[74,76]

### D. Temporal variation of drag and lift coefficients

Based on the spatiotemporal wall pressure information forecast, the force coefficient information is further investigated, including the lift coefficient $C_l$ and the drag coefficient $C_d$. Similarly to Sec. V B, the standard deviation of the lift coefficient $C_{l,std}$ and the mean value of the drag coefficient $\overline{C}_d$ are compared between the forecast results and the experimental data. The force coefficients are calculated from Eqs. (21) and (22). Figure 10 presents the aerodynamic coefficients between the forecast results and experimental data from the statistical perspective. It can also be derived that the forecast information shares mean drag coefficient information ($\overline{C}_d$) close to the experimental data, and the standard deviation of the lift coefficient ($C_{l,std}$) varies slightly.

Figure 11 compares the instantaneous force coefficient, including the drag coefficient $C_d$ and the lift coefficient $C_l$ in the temporal domain. The definition of the x axis is the dimensionless time, which is the same as in Fig. 7. As shown in Fig. 11(a), the temporal forecast of drag coefficient $C_d$ shows the value range $C_d \in (1.55, 1.72)$ , which is slightly narrower than the drag coefficient of experimental data as $C_d \in (1.50, 1.85)$. The drag coefficient of the experimental data presents a larger fluctuation range. A similar trend is detected in the forecast temporal development of the lift coefficient shown in Fig. 11(b). The forecast temporal lift coefficient $C_l \in (-0.76, 0.75)$, which is slightly narrower than the experimental lift coefficient range as $C_l \in (-1.1, 1.01)$. Therefore, to better understand the discrepancy between the forecast result and the experimental data and how this discrepancy arises, we will investigate the spatial distributions of instantaneous wall pressures further in Sec. V E.

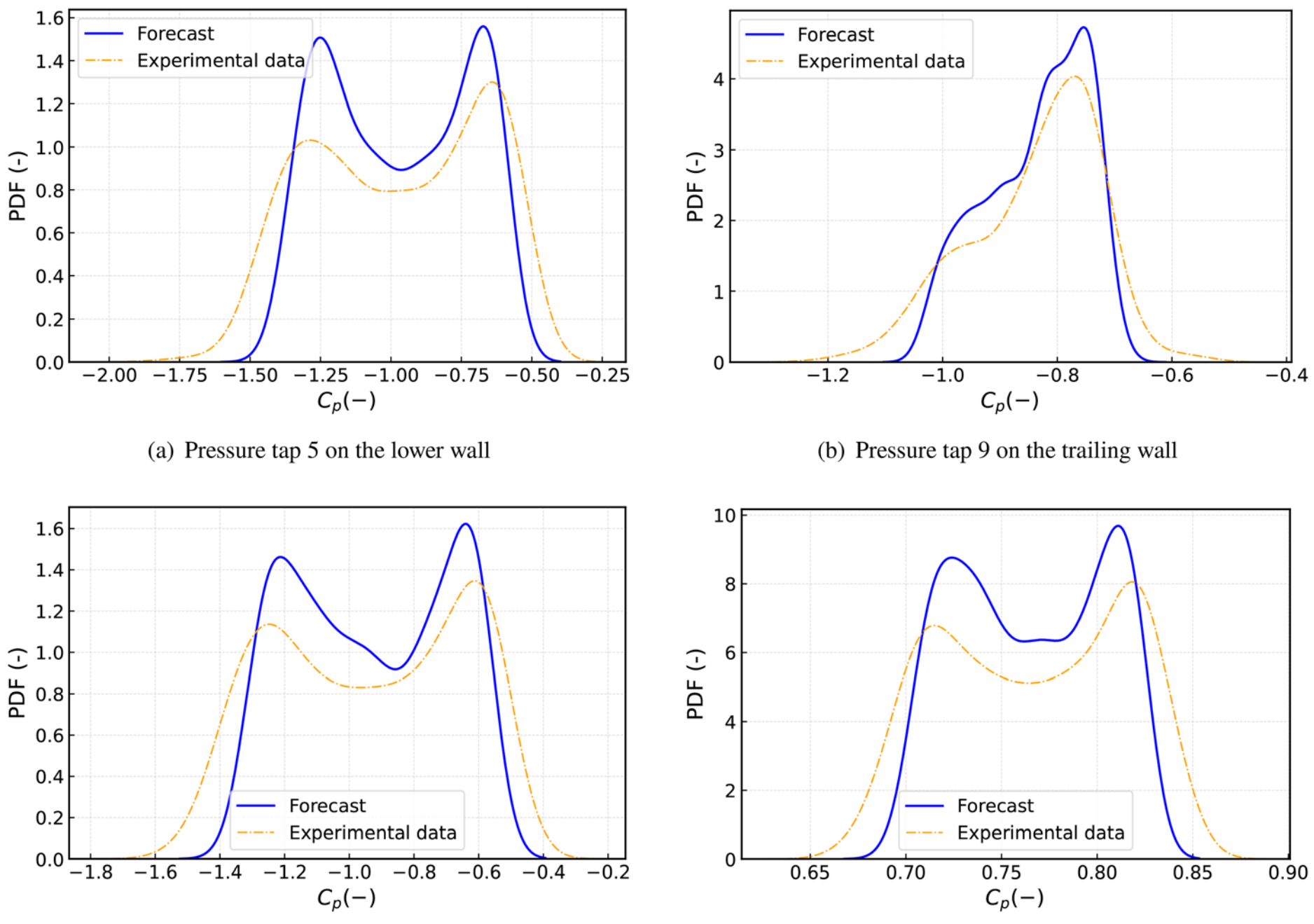

(a) Pressure tap 5 on the lower wall
(b) Pressure tap 9 on the trailing wall
(c) Pressure tap 18 on the upper wall
(d) Pressure tap 22 on the leading wall

**FIG. 8.** Wall pressure temporal information statistical comparison between forecast result and experimental data at four different spatial pressure taps. In y axis, PDF denotes the probability density function. In both x and y axis, the sign (-) denotes the dimensionless value.

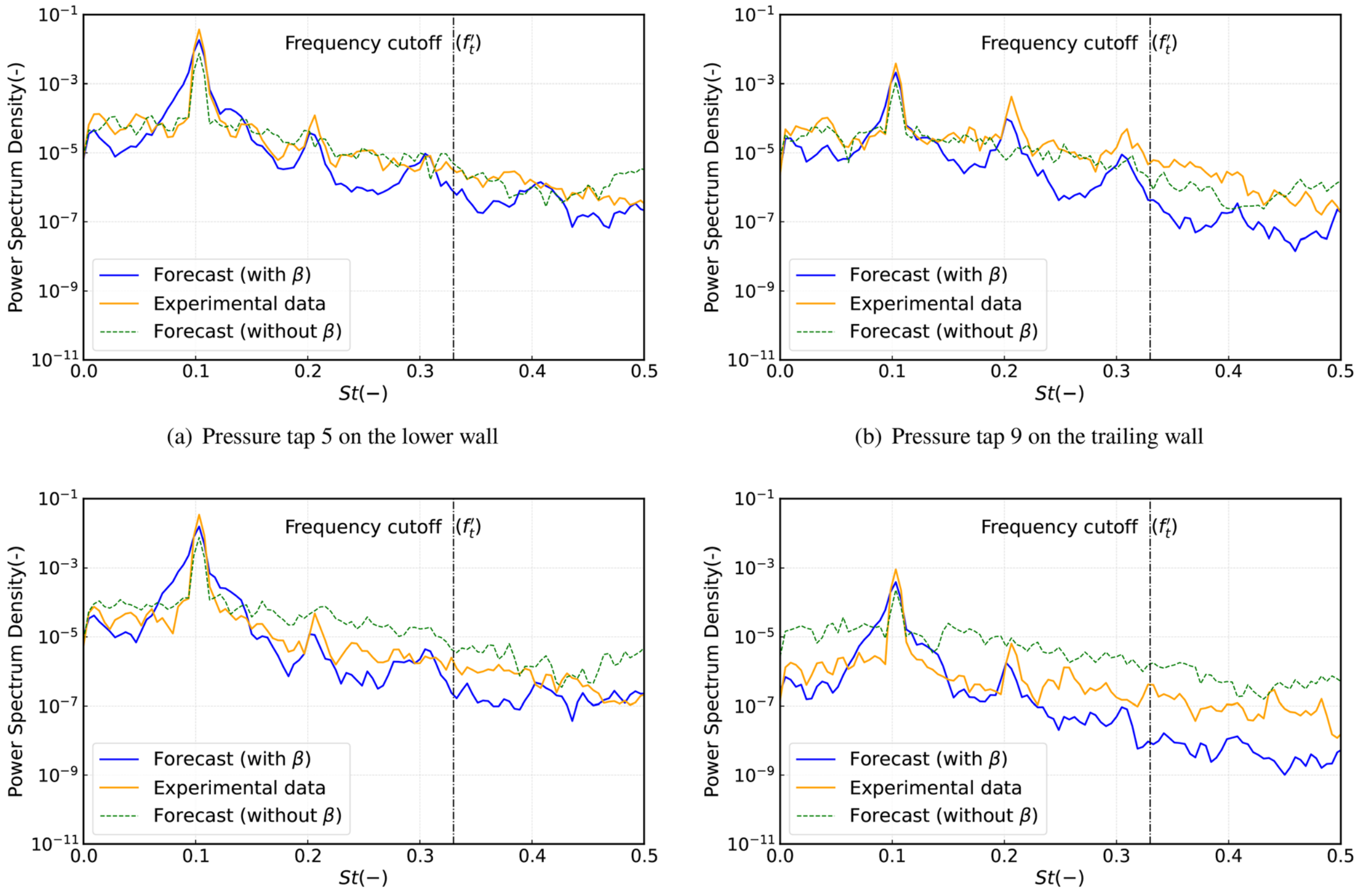

(a) Pressure tap 5 on the lower wall
(b) Pressure tap 9 on the trailing wall
(c) Pressure tap 18 on the upper wall
(d) Pressure tap 22 on the leading wall

**FIG. 9.** Forecast wall pressure fluctuating information power spectrum density (PSD) at different spatial pressure taps. In x axis, the *St* denotes the calculated Strouhal number based on Eq. (24). In both x and y axis, the sign (-) denotes the dimensionless value.

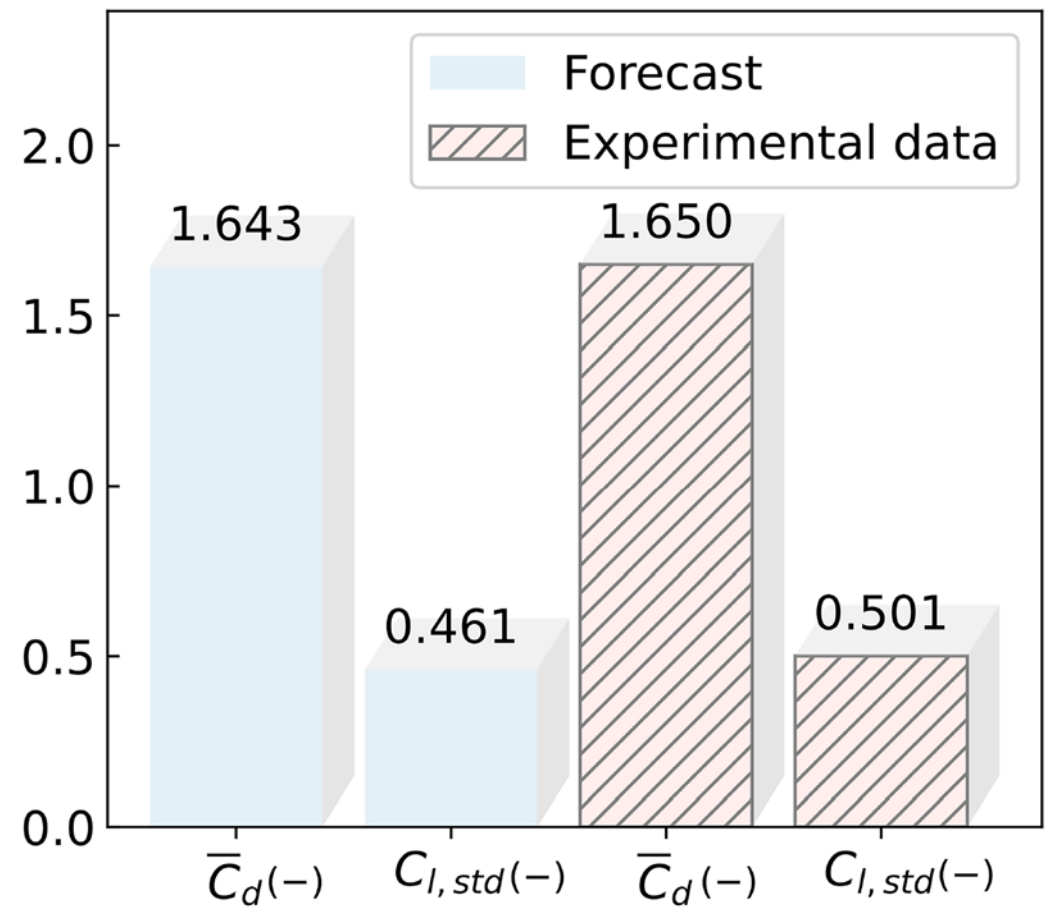

**FIG. 10.** Aerodynamic statistics ($\overline{C}_d$, $C_{l,std}$) comparison between forecast result and experimental data. In x axis, the sign (-) behind each terminology denotes the dimensionless value.

## E. Instantaneous spatial pressure distribution

As indicated by the temporal history diagram of the lift coefficient $C_l$ shown in Fig. 11(b), several instantaneous temporal snapshots are selected to understand the instantaneous pressure coefficient distribution in the spatial domain. In total, five snapshots are selected to compare instantaneous pressure information at various spatial points as presented in Fig. 12, including $\Delta T/\delta t = 355, 358, 367, 983$, and 989. These five instantaneous temporal snapshots cover the lift coefficient $C_l$ residing in the peak value (snapshot 989), valley values (snapshot 355, 367, 983), and the large discrepancies (snapshot 358) between the forecast result and the experiment data.

Figure 13 compares the experimental data and the forecast result. It can be seen that when the instantaneous lift coefficient $C_l$ is located in the peak ($\Delta T/\delta t = 989$) and valleys ($\Delta T/\delta t = 355, 367$ and 989), the instantaneous spatial pressure distribution of the forecast results agrees well with the experimental data shown in Figs. 13(a) and 13(c)–13(e), including the upper and lower rectangular cylinder walls. These agreements between the forecast results and the experimental data demonstrate the strong capability of the DeepUFNet model to forecast spatiotemporal wall pressure information. In addition, for snapshot $\Delta T/\delta t = 358$ shown in Fig. 13(b), the discrepancy lies mainly in the upper wall and lower wall as pressure tap Nos. 1–8 and 14–21. This phenomenon is triggered by the complex fluid flow around the upper and lower walls of flow around rectangular cylinders.[23,46]

## F. Spatiotemporal correlation

In this part, based on the wall pressure information forecast, the spatiotemporal pressure correlation on the upper wall is investigated to estimate the model performance in understanding physics. Taking two pressure taps on the upper wall of the rectangular cylinder as an example, the spatiotemporal correlation of two different pressure taps in different spatial domains $x$ and $x + i\Delta x$ is calculated from the spatiotemporal pressure distribution.[20,92–94] The spatiotemporal correlation coefficient can be calculated from the wind tunnel experimental data with the following equation:

$$R(i\Delta x, j\Delta t) = \frac{\overline{p(x,t)p(x+i\Delta x, t+j\Delta t)}}{\sqrt{\overline{p(x,t)^2}\,\overline{p(x+i\Delta x, t+j\Delta t)^2}}}, \tag{29}$$

where $p(x, t)$ is the wall pressure information at the spatial point $x$ at time $t$, $p(x + i\Delta x, t + j\Delta t)$ is the wall pressure information at the spatial point $x + i\Delta x$ at time $t + j\Delta t$, and $\overline{()}$ means the average operator in spatial variable $x$ and time $t$. From Eq. (29), there are two types of pressure information in different spatial points and different temporal intervals. $i\Delta x$ means the spatial separation between two pressure taps and $j\Delta t$ means the temporal separation. In the calculation of Eq. (29), there are eight spatial points in the upper wall as shown in Fig. 3. Therefore, the maximum space interval is set as $i_{max}\Delta x = 6\Delta x$, where $\Delta x$ denotes a single space interval as depicted in Fig. 3. For the temporal interval, there are a total of 1000 snapshots for the forecast result, and the maximum temporal interval is set as $j_{max}\Delta t = 700\Delta t$. For example, $R(2\Delta x, 5\Delta t)$ denotes the pressure correlation coefficient for two pressure taps with the $2\Delta x$ spatial interval and the $5\Delta t$ temporal interval. The results of the correlation coefficient for both the forecast wall pressure coefficient and the experimental data within a short period ($7\,\Delta t$) are shown in Fig. 14. In the figure, the z axis $R$ denotes the correlation coefficient for a particular spatiotemporal separation. The x axis is the dimensionless spatial interval index $i$, and the y axis denotes the temporal interval index $j$, which indicates the spatial and temporal separations as $i\Delta x$ and $j\Delta t$, respectively. Here, $\Delta t = 1/f_s$ is the data sampling interval in the physical time domain, which is the same unit as $\delta t$ shown in Figs. 7 and 11. From the correlation results of the experimental data shown in Fig. 14(b), taking a fixed time interval $1\Delta t$ as an example, when the space interval increases, the correlation coefficient decreases, and this trend is also observed in the correlation coefficient of the forecast results shown in Fig. 14(a). The correlation value decrement phenomenon at a fixed time interval corresponds to others' work.[95–98]

Specifically, taking a longer temporal period ($25\,\Delta t$) as the length of the investigation, correlation coefficients of different spatial separation are studied. From the results shown in Fig. 15, it can be seen that the correlation coefficient ($R$) of the forecast result has peaks identical to those of the experimental data. The valley values are different, and the correlation coefficients of the forecast data around the valleys are larger than the correlation coefficients of the experimental data. In other words, the DeepUFNet model has the ability to make a higher correlation forecast on the spatiotemporal wall pressure information. Some spatiotemporal information with lower correlations is overestimated by the DeepUFNet model shown in Fig. 14. For example, when the temporal interval $j\Delta t = 5\Delta t$, the correlation of the forecast wall pressure in different spatial intervals is higher than that of the experimental data. In addition, the DeepUFNet model detects the shift of the peak in the spatiotemporal domain, for example, the peak is located at $j\Delta t = 12\Delta t$ with $i\Delta x = \Delta x$ shown in Fig. 15(a), and the peak shifts to $j\Delta t = 10\Delta t$ in $i\Delta x = 6\Delta x$ shown in Fig. 15(f). This shift is induced by the downstream propagation of wall pressure on the upper surface, and similar downstream shifting is also presented in existing work, such as turbulent boundary layer,[95,99,100] turbulent channel flow,[101,102] and turbulent shear flow.[97,103] Interestingly, from the correlation coefficient results shown in Fig. 15, the periodicity timescale ($\mathcal{T}$) of the correlation coefficient is around $\mathcal{T} = 11 \sim 12\Delta t$ calculated from peaks or valleys. Based on the periodicity timescale of the correlation coefficient $\mathcal{T}$, the oscillation frequency $f_R$ of the correlation coefficient can be calculated as $f_R = 1/\mathcal{T} \approx f_s/11$, where $f_s = 400$ Hz is the pressure scanning frequency presented in Sec. IV. It is found that the correlation coefficient oscillation frequency in Fig. 15 ($f_R$) corresponds to the rectangular wake vortex shedding dominant frequency ($f$) as shown in Eq. (24) (i.e., $f_R = f$), while such periodicity is absent in turbulent wall-bounded shear flows without wake.[93,104]

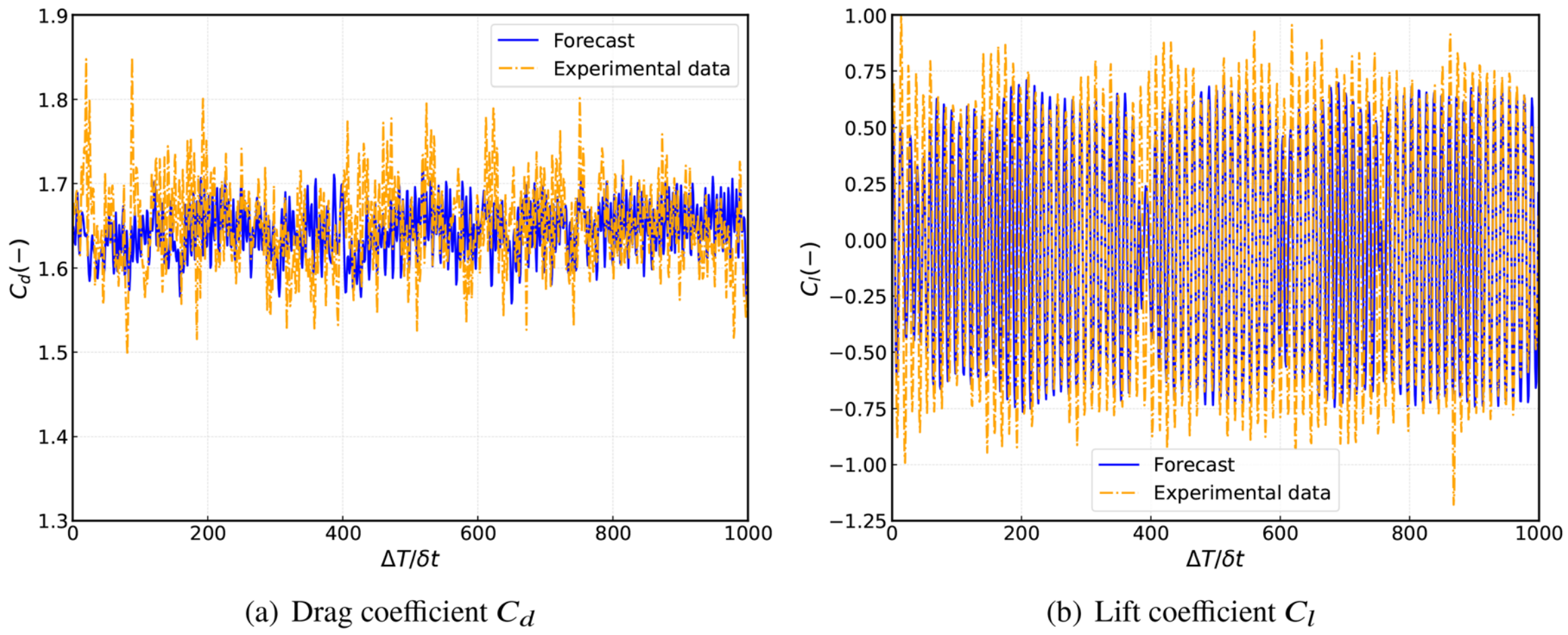

**FIG. 11.** Temporal history instantaneous force coefficient comparison between the forecast result and experimental data. For both figures, in y axis, the sign (-) denotes dimensionless values of drag coefficient ($C_d$) and lift coefficient ($C_l$). In x axis, $\Delta T$ denotes the physical time information accumulating from the start of forecast information with unit *s*, and $\delta t$ denotes the pressure scanning interval with unit *s*.

## G. Spectral proper orthogonal decomposition

In this subsection, the Spectral Proper orthogonal decomposition (SPOD) technique is applied to investigate the forecast result and experimental data.[105–107] Packages based on Python language are used in this work for the implementation of SPOD.[108,109] We study the mode information and the frequency information from the perspective of the energy distribution and reconstruct the pressure information from the SPOD analysis. As shown in Fig. 16, the SPOD mode-frequency energy distribution of both the forecast result and the experimental data look similar. In Fig. 16, the first ten modes are visualized in different colors, and other modes with relatively low energy are in gray color. As shown in both Fig. 16(a) of the forecast result, before the frequency cutoff, there are three frequency peaks, and these peaks are also observed in Fig. 16(b) of the experimental data. However, these peaks have different bandwidths as shown in the forecast result and in the experimental data. The three frequency peak bandwidths are border in the forecast result as illustrated in Fig. 16(a), but the peaks bandwidths are narrower in the experimental data shown as sharp peaks as presented in Fig. 16(b). Similar peak bandwidth differences are observed in Fig. 9, as sharp peaks are located in the dominant frequency zone in the experimental data. In other words, in the experimental data, the energy distribution concentrates within a narrow frequency zone.

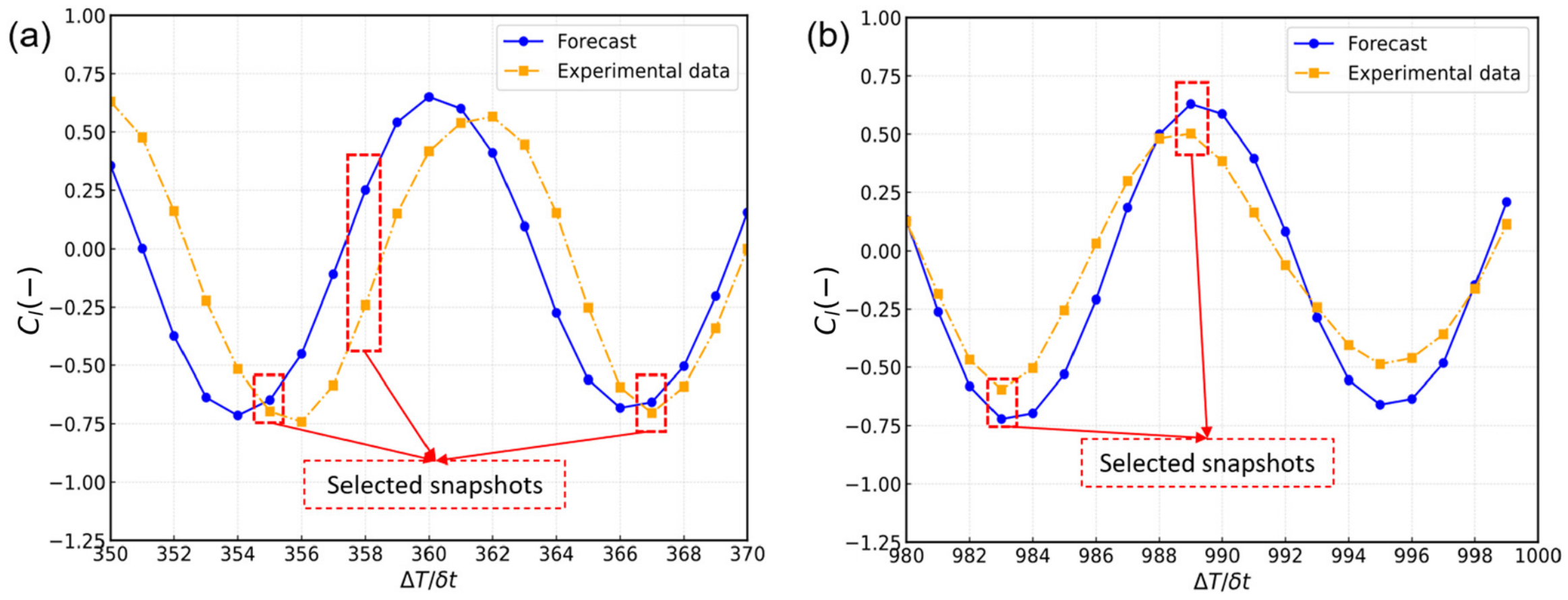

**FIG. 12.** Instantaneous snapshot selection to understand the wall pressure spatial distribution based on the lift coefficient $C_l$ temporal history. In total, five snapshots are selected, including (a) two snapshot with $C_l$ at the valley, and the other one snapshot $C_l$ has large discrepancy between experimental data and forecast result, (b) one snapshot with $C_l$ at the peak, and one snapshot with $C_l$ at the valley. In y axis, the sign (-) denotes the dimensionless value. In x axis, $\Delta T$ denotes the physical time information accumulating from the start of forecast information with unit *s*, and $\delta t$ denotes the pressure scanning interval with unit *s*.

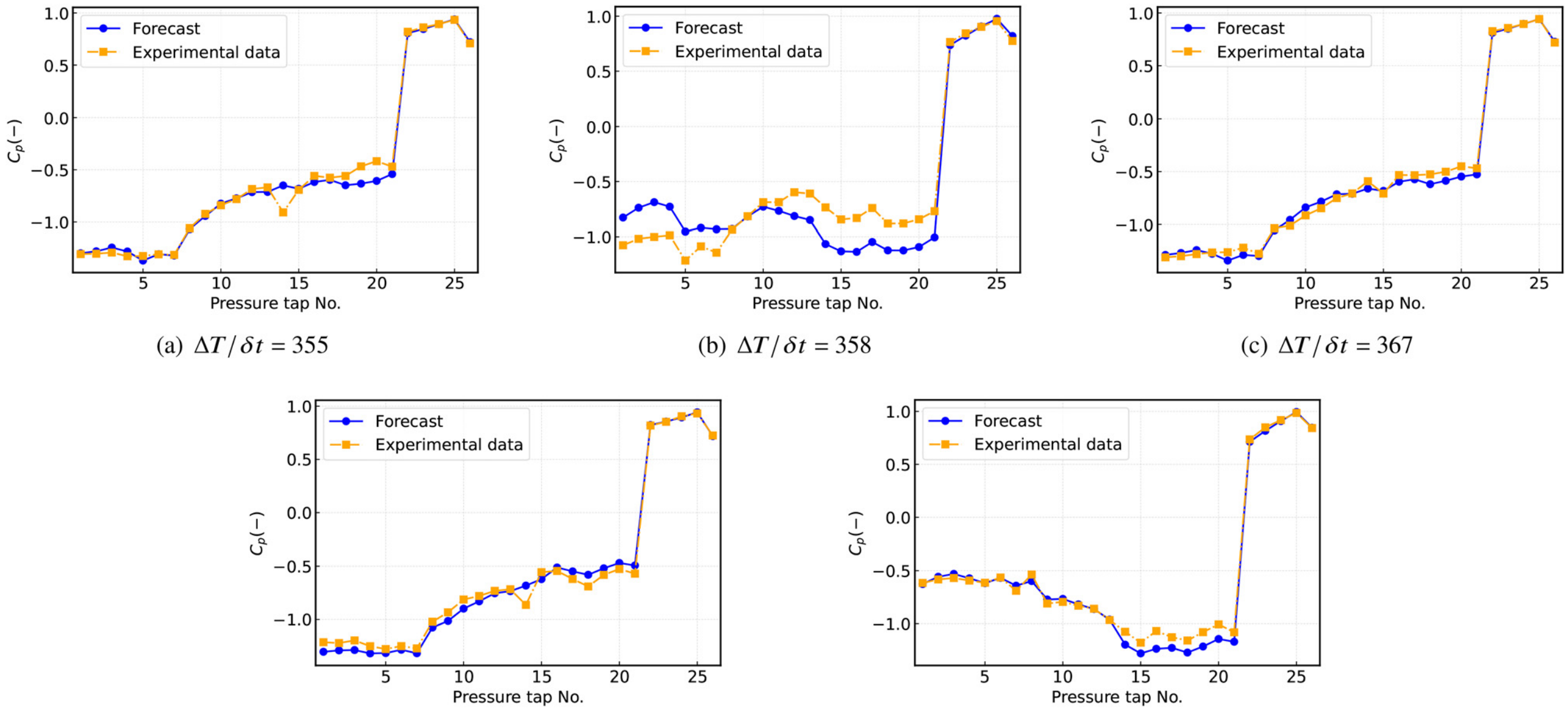

(a) $\Delta T/\delta t = 355$ (b) $\Delta T/\delta t = 358$ (c) $\Delta T/\delta t = 367$

(d) $\Delta T/\delta t = 983$ (e) $\Delta T/\delta t = 989$

**FIG. 13.** Spatial pressure coefficient ($C_p$) distribution at different instantaneous snapshots. In y axis, the sign (-) denotes the dimensionless values of pressure coefficient. In x axis, $\Delta T$ denotes the physical time information accumulating from the start of forecast information with unit *s*, and $\delta t$ denotes the pressure scanning interval with unit *s*.

Furthermore, we calculate the accumulative energy distribution on each particular frequency and reconstruct wall pressure based on some particular modes and frequencies as shown in Fig. 17. From the energy distribution in the frequency domain presented in Fig. 17(a), we can find that, for both the experimental data and the forecast result, the dominant frequency takes most of the energy from the perspective of the frequency domain. However, a discrepancy is observed in the bandwidth for the first peak. In addition, on the basis of the SPOD results, the wall pressure coefficient is reconstructed from the frequency and mode domain. We reconstruct the wall pressure coefficient based on mode 1 of all frequencies. The reconstructed wall pressure coefficient for both the forecast results and the experimental data are close to the temporally averaged wall pressure coefficient $\overline{C}_p$ as shown in Fig. 17(b). This behavior is similar to some existing results, which have also accurately reconstructed the temporally averaged characteristics[110,111] based on mode 1 of all frequencies.

## VI. MODEL EXTRAPOLATION: FULL SPATIAL DIMENSION RECONSTRUCTION WITH SPARSE INPUT

From the above results, it can be seen that the DeepUFNet model developed in this work can forecast the spatiotemporal wall pressure information with full spatial dimension input information $\mathcal{P}_{in}$ of size (1000, 26), where 1000 is the temporal snapshots and 26 is the number of spatial points. However, as discussed in the introduction, in some practical scenarios and engineering practices, full spatial dimension information is not always available or easily accessible, and the collected data are accompanied by scarcity or noise. Inspired by some literature that reconstructs full-scale information with deep learning techniques based on sparse representation,[112–119] we further evaluated the potential of the DeepUFNet model for extrapolation, i.e., forecasting spatiotemporal wall pressure information with limited information. We consider a practical scenario in which only partial spatial observations, that is, sparse spatial information, are available in the input. Motivated by the quantitative investigation of spatiotemporal correlations in Sec. V F, there is a high spatiotemporal correlation for the eight pressure taps on the upper wall. In symmetry, there is high spatiotemporal correlation on eight pressure taps of the lower wall. This motivates us to reduce the number of spatial pressure taps on both the upper and lower walls used in the DeepUNFet model. This sparse spatial dimension information is intended to assess the spatial extrapolation capability of DeepUFNet when there are sparse spatial measurement locations, which is widely seen in engineering practices. This setting also simulates practical engineering conditions, such as sparse sensor deployment or data corruption and evaluates the model's generalization beyond the observed spatial domain.

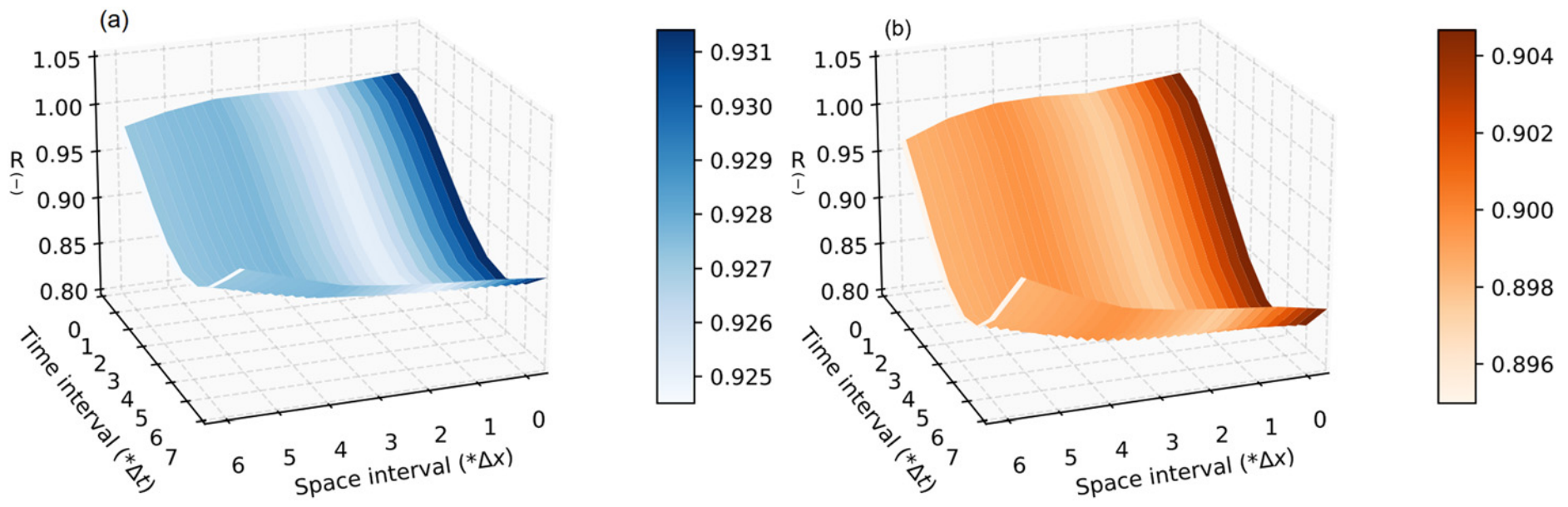

**FIG. 14.** Correlation coefficient comparison between forecast result and experiment data with different spatial intervals. (a) Forecast wall pressure spatiotemporal correlation distribution. (b) Experimental data spatiotemporal correlation distribution. The time interval index axis is showing index *j* in Eq. (29) leading to temporal separation as $j\Delta t$. The spatial interval axis is displaying index *i* in Eq. (29) corresponding to the spatial separation of the pressure tap distance on the rectangular cylinder wall as $i\Delta x$. In the vertical axis, the sign (-) denotes the dimensionless value.

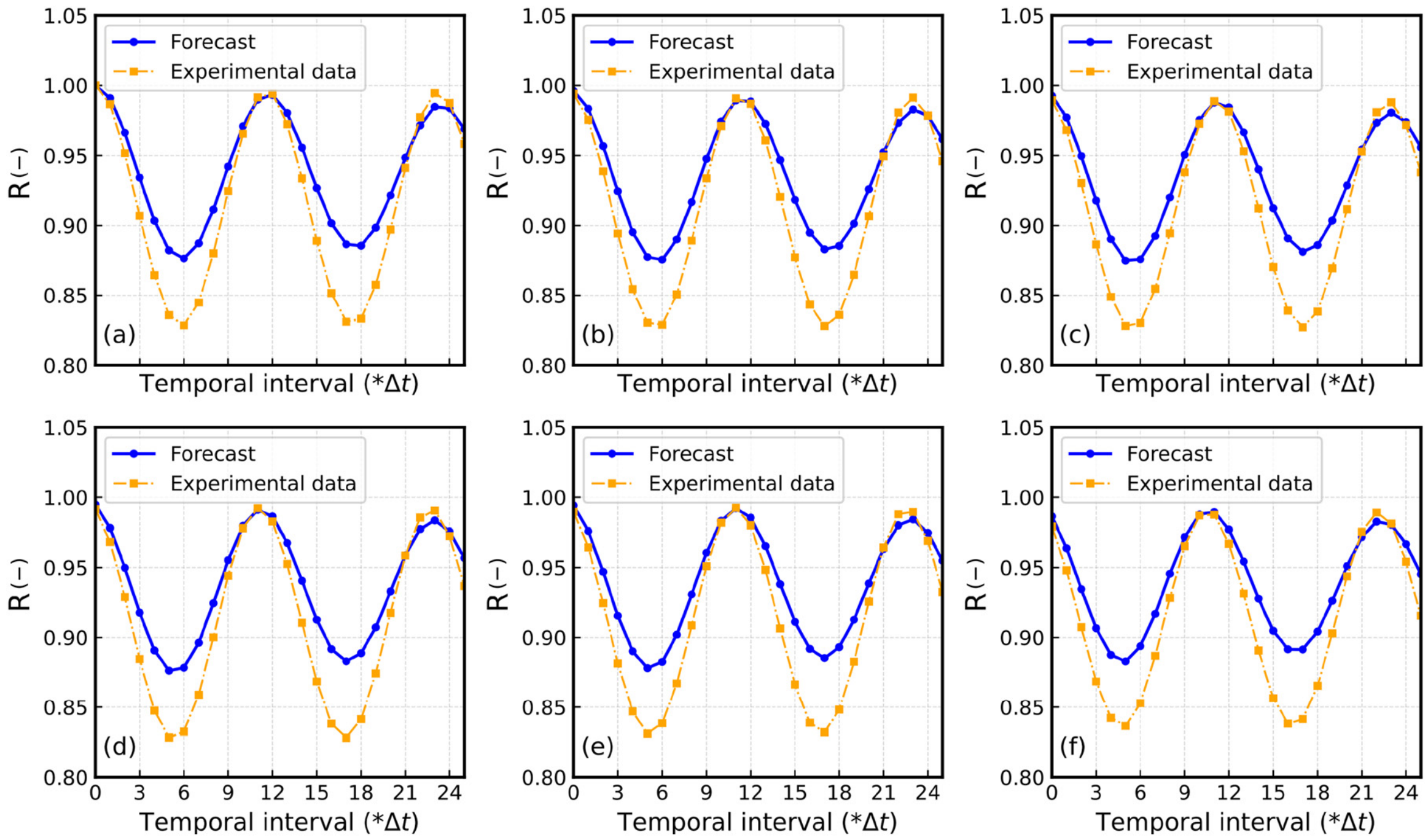

FIG. 15. Correlation coefficient comparison between forecast result and experiment data with different spatial intervals. (a) $\Delta x$, (b) $2\Delta x$, (c) $3\Delta x$, (d) $4\Delta x$, (e) $5\Delta x$, (f) $6\Delta x$. The x axis in each figure denotes the temporal interval index $j$ in Eq. (29) corresponding to temporal separation as $j\Delta t$. In y axis, the sign (-) denotes the dimensionless value.

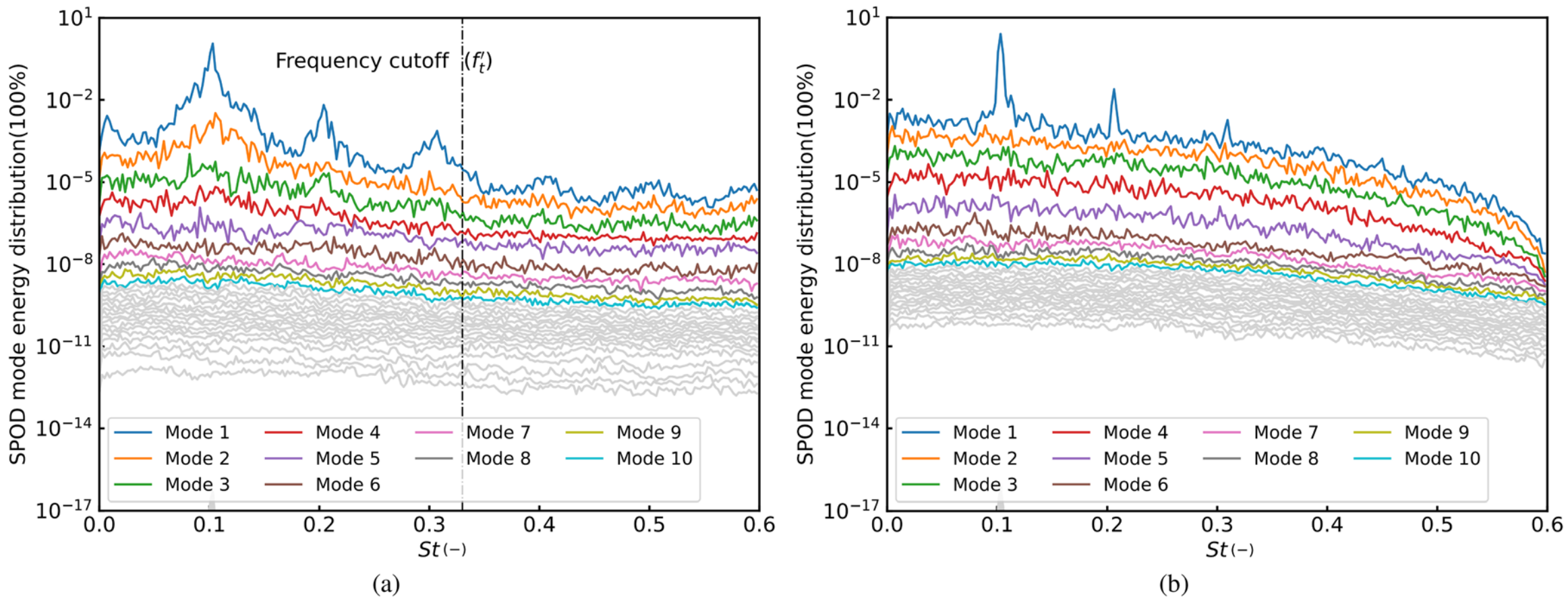

FIG. 16. SPOD mode-frequency energy distribution. (a) The SPOD mode-frequency results calculated from the forecast wall pressure information, (b) The SPOD mode-frequency results calculated from the experimental data. The x axis is the Strouhal number calculated from Eq. (24), and the sign (-) denotes the dimensionless value. First 10 modes are visualized in colors.

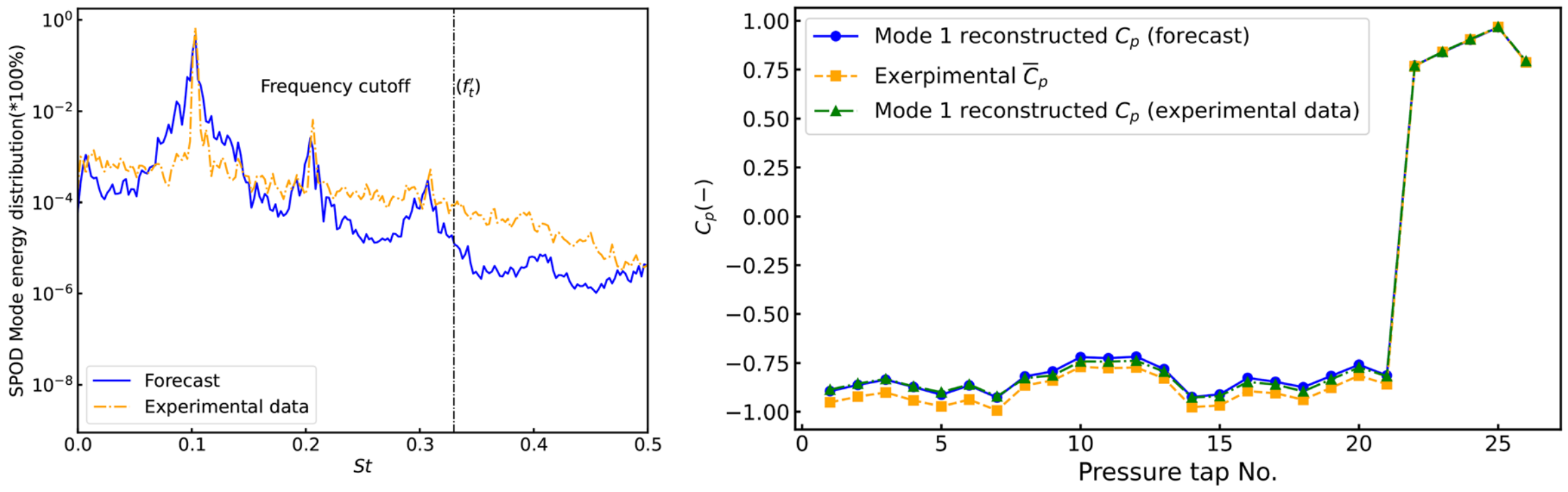

**FIG. 17.** (a) SPOD frequency energy distribution indicating the energy distribution in frequency domain with all modes. (b) The reconstructed wall pressure coefficient with mode 1 of all frequencies, and the sign (-) in y axis denotes the dimensionless value.

Mathematically, instead of using the full spatial resolution with spatial points $|\mathbf{\Omega}| = 26$, we define a reduced spatial domain $\mathbf{\Omega}_1 \subset \mathbf{\Omega}$ with $|\mathbf{\Omega}_1| < |\mathbf{\Omega}|$ and provide the model with pressure data $\mathcal{P}_{in}(\mathbf{\Omega}_1, \mathbf{\Theta})$ of shape $(1000, k)$ where $k = 18$ in this extrapolation test. Specifically, the spatial pressure taps on the upper wall and the lower wall are reduced to half, and the total of 18 spatial pressure taps and the pressure tap index are shown in Fig. 18. From the figure, it can be seen that the new pressure taps used in both the upper and lower walls ($\mathbf{\Omega}_1$) are parts of the original total pressure taps ($\mathbf{\Omega}$) shown in Fig. 3. In this setup, the DeepUFNet model is trained and evaluated to perform the following spatiotemporal extrapolation task:

$$\mathcal{G}_\theta[\mathcal{P}_{in}(\mathbf{\Omega}_1, \mathbf{\Theta})] = \mathcal{P}_{out}(\mathbf{\Omega}, \mathbf{\Theta} + \Delta T), \tag{30}$$

which corresponds to forecasting future wall pressure fields in a higher resolution spatial domain $\mathbf{\Omega}$ based on fewer spatial sensors in $\mathbf{\Omega}_1$ and prior time series observations.

In the implementation stage, we reduced the number of spatial pressure taps on both the upper and lower walls. The number of spatial points is reduced from 26 to 18 in the input part, and the output information $\mathcal{P}_{out}$ is kept the same; that is, the input information size is (1000, 18), and the output size is (1000, 26). The new spatial difference in the upper and lower wall is 10 mm, while the spatial difference in the leading and trailing wall remains the same as 5 mm. The target is to forecast future spatiotemporal wall pressure information using sparse wall pressure information.

Based on the information and objective described above, the model is trained on the same database but with different spatial dimensions in the input part. The model architecture keeps the same apart from the output layer with a dimension increment. The hyperparameters of the model and the training settings remain the same as shown in Table I. The physical high-frequency loss control coefficient $\beta$ is embedded into the model training stage for all results in this section. We will then evaluate the model performance to forecast spatiotemporal variation of wall pressure.

Similar to Sec. V B, in the estimation of the model extrapolation ability, the temporally averaged wall pressure coefficient ($\overline{C}_p$) and the standard deviation value ($\sigma$) of the forecast wall pressure coefficients at different spatial points with a sparse spatial input are studied. The results ($\overline{C}_p$, and $\sigma$) are compared with the experimental data and the forecast wall pressure with the full spatial input. As shown in Fig. 19, with the input of a sparse spatial dimension, the temporally averaged wall pressure coefficients ($\overline{C}_p$) on the upper and lower walls, i.e., pressure tap Nos. 1–8 and 14–21, present a minor difference compared with the temporally averaged wall pressure coefficients of experimental data and temporal average of the forecast wall pressure coefficient with full spatial input. In the leading and trailing walls associated with pressure tap Nos. 9–13 and 22–26, the temporally averaged wall pressure coefficients ($\overline{C}_p$) of the forecast wall pressure with the sparse spatial input information remains the same as both experimental data and the forecast results with full spatial input, as demonstrated in Fig. 19. The above finding suggests that reducing the input spatial dimension on both the upper and lower walls can still lead to high accuracy in the forecast of temporally averaged wall pressure in full spatial dimensions.

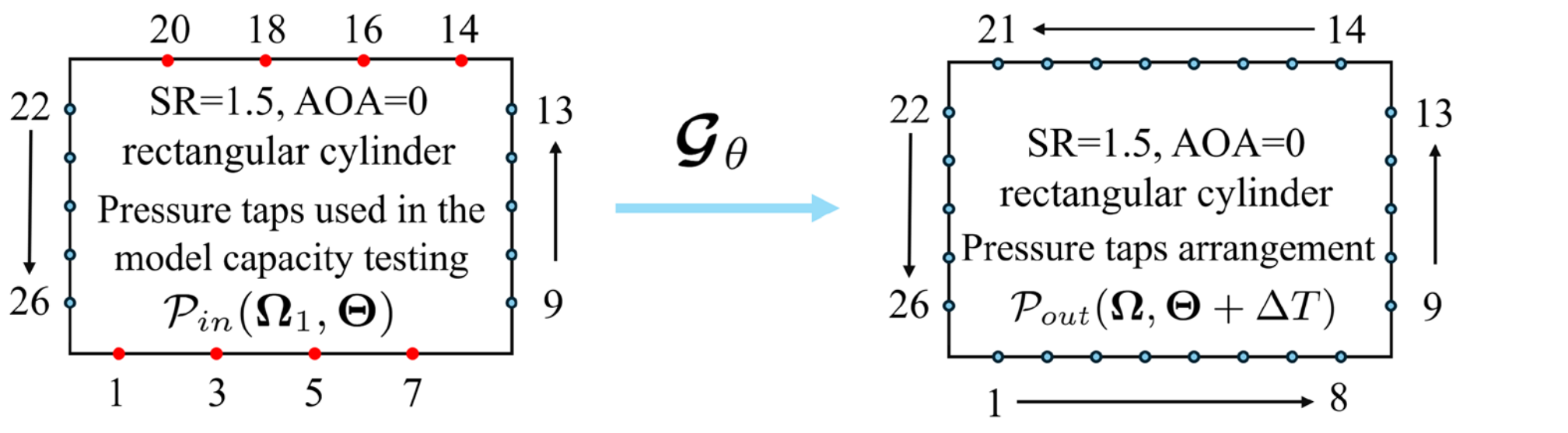

**FIG. 18.** Sparse spatial pressure taps selection to test model extrapolation capacity: half number of the upper and lower pressure taps are used, and the red color denotes the used pressure taps on both walls. The updated spatial difference on upper and lower wall is 10 mm. The target is to project the $\mathcal{P}_{in}(\mathbf{\Omega}_1, \mathbf{\Theta})$ to $\mathcal{P}_{out}(\mathbf{\Omega}, \mathbf{\Theta} + \Delta T)$ with the DeepUFNet model $\mathcal{G}_\theta$.

The standard deviation value ($\sigma$) of the wall pressure has some variations indicated by the results shown in Fig. 19. The standard deviation value of the forecast wall pressure on upper and lower walls with sparse spatial input is smaller than those of the experimental data and the forecast result with full spatial input. The comparison between these three types of results, including forecast result with sparse spatial input, forecast results with full spatial input, and experimental data, indicates that inputting sparse spatial information to the DeepUFNet model will affect the model performance. This phenomenon can be further understood from the perspective of flow dynamics around the rectangular cylinder: the flow presents flow separation in the leading wall corner, and the separated flow generates complex vortex near the upper and lower walls, which also interacts with the vortex in the wake.[23] However, in the near trailing wall zone, there is no obvious vortex structures visualized from flow field.[22,46] On this condition, the complex flow near both the upper and lower walls leads to a complex pressure distribution on the upper and lower walls. When reducing the spatial dimension to DeepUFNet, the DeepUFNet model understands less information from the sparse spatial input, such as missing information on small flow structures within the spatial resolution, as well as the fluctuation of the wall pressure caused by the small flow structure. Therefore, when forecasting the full spatial dimension information with sparse spatial input, some small flow structures induced wall pressure fluctuations within the spatial interval are smoothly proceeded shown as lower temporally averaged pressure coefficient and lower standard deviations on the upper and lower walls in Fig. 19. Amazingly, the difference in temporal average values of the pressure coefficient is very small between the forecast wall pressure with sparse spatial input and the forecast wall pressure with full spatial input.

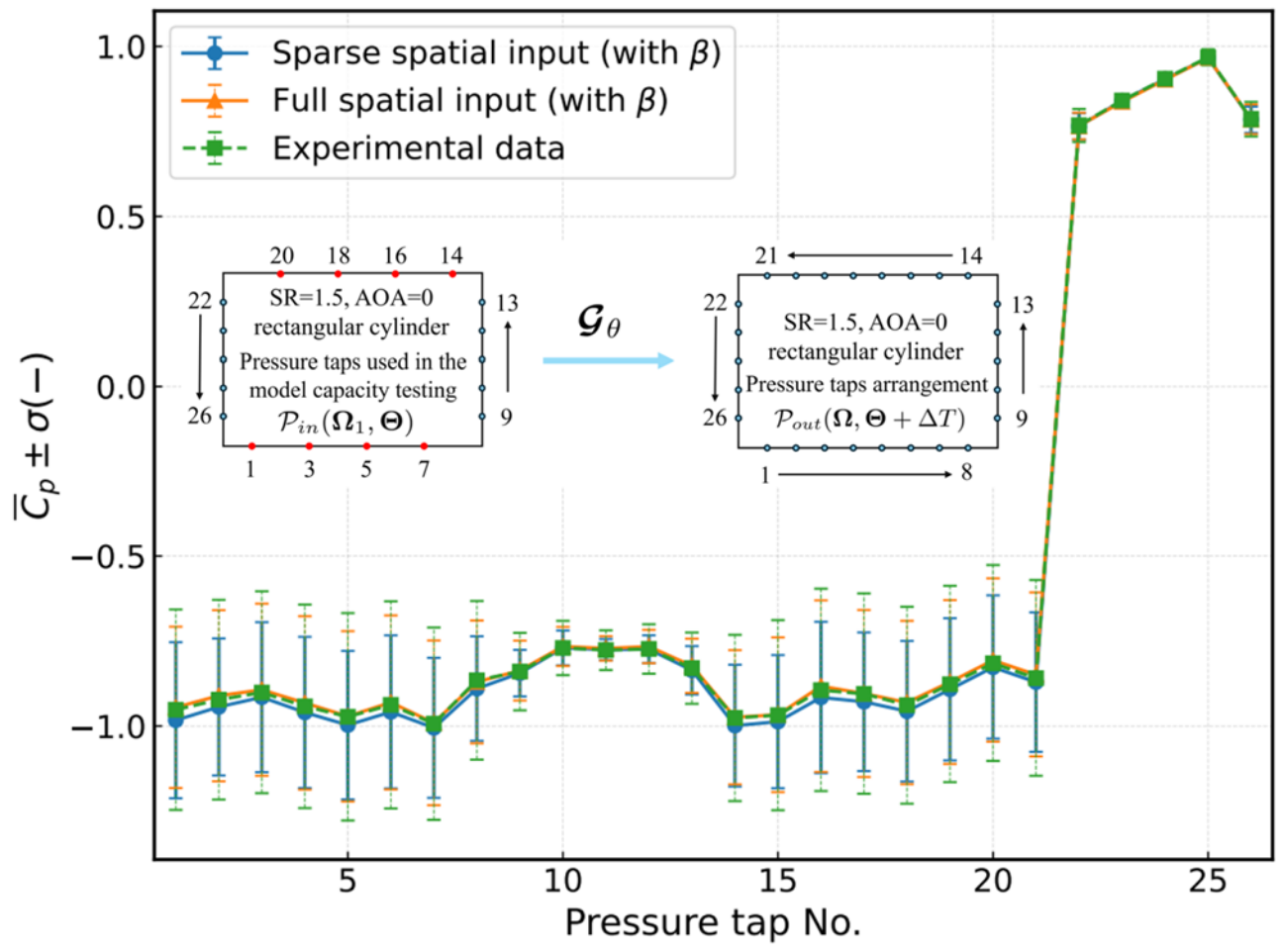

**FIG. 19.** Comparison of temporally averaged wall pressure coefficient ($\overline{C}_p$) and the standard deviation value ($\sigma$) of wall pressure at different spatial points from different spatial input. Different scatter points shapes denote the different types of temporally averaged wall pressure coefficient $\overline{C}_p$, and the vertical bars denote standard deviation values. In y axis, the sign (-) denotes the dimensionless values of $C_p \pm \sigma$.

In addition, as expressed in Sec. V C, the power spectrum density of fluctuating pressure coefficients at four spatial points, that is, pressure tap Nos. 5, 9, 18, 22, on various walls, is studied to compare the fluctuating characteristics with the forecast results with full spatial information as inputs. Here, we select these four spatial points again for a direct comparison to demonstrate the model performance with sparse spatial information input. Figure 20 shows the fluctuating power spectrum density of three types of data at these four spatial points, including the forecast wall pressure with sparse spatial input, the forecast wall pressure with full spatial input, and the experimental data. As shown in Fig. 20, with either the sparse spatial input or the full spatial input as well as embedding the high-frequency loss control coefficient $\beta$ in the model training, the DeepUFNet model can capture the general fluctuating characteristics. Three main peaks ($St_m \approx 0.103$, $2St_m \approx 0.206$, and $3St_m \approx 0.31$) are observed in all four figures corresponding to four spatial measurement positions on various walls. In addition, the power spectrum density of the forecast wall pressure with sparse spatial input and the power spectrum density of the forecast wall pressure with full spatial input are close to each other. Though missing small flow structure-induced wall pressure fluctuation information in the spatial domain, the DeepUFNet model can still reconstruct the wall pressure high-frequency fluctuation by sparse spatial pressure measurement with high-frequency loss control coefficient $\beta$ embedding as shown in Fig. 20.

The above finding indicates that using sparse spatial information as input with the physical frequency loss control coefficient $\beta$ in the model stage can forecast the wall pressure with satisfactory results. In addition, reducing the spatial measurement positions in the upper and lower walls as the input information will not significantly deteriorate the model performance, which provides a bright perspective on addressing engineering problems, such as high-accuracy sparse reconstruction in the circumstance of data missing.

Finally, five spatial distributions of instantaneous wall pressure forecast by the DeepUFNet model with sparse spatial input are presented to compare with two other types of data, including experimental data and spatial distributions of instantaneous wall pressure predicted by full spatial input. From the comparison results shown in Fig. 21, it can be seen that the instantaneous forecast of wall pressure distribution with sparse spatial information input is close to the experiment data in four snapshots, including $\Delta T/\delta t = 355, 367, 983$, and 989. Some minor discrepancy is observed in snapshot $\Delta T/\delta t = 358$ in pressure tap Nos. 12–15. As shown in Fig. 21(b), the forecast spatial distribution of instantaneous wall pressure with sparse spatial input is closer to the experimental data in contrast with the instantaneous forecast wall pressure with full spatial input. Interestingly, shown in Fig. 21(b), the instantaneous forecast wall pressure with full spatial input has a slightly larger difference compared to the experimental data on the upper and lower walls, i.e., pressure tap Nos. 1–8 and 14–21. This phenomenon clearly presents the model extrapolation abilities, particularly on the upper and lower walls with sparse spatial information available.

From the above investigation of model spatial extrapolation capability, including statistical information analysis, fluctuating power spectrum density analysis, and spatial distribution of instantaneous wall pressure coefficients, it can be concluded that reducing the spatial input dimension to sparse information has negligible negative effects on the spatiotemporal wall pressure forecast apart from the slight difference in the wall pressure standard deviation as shown in Fig. 19. The model extrapolation ability test indicates that the DeepUFNet model developed in this work is robust in the spatiotemporal wall pressure forecast task.

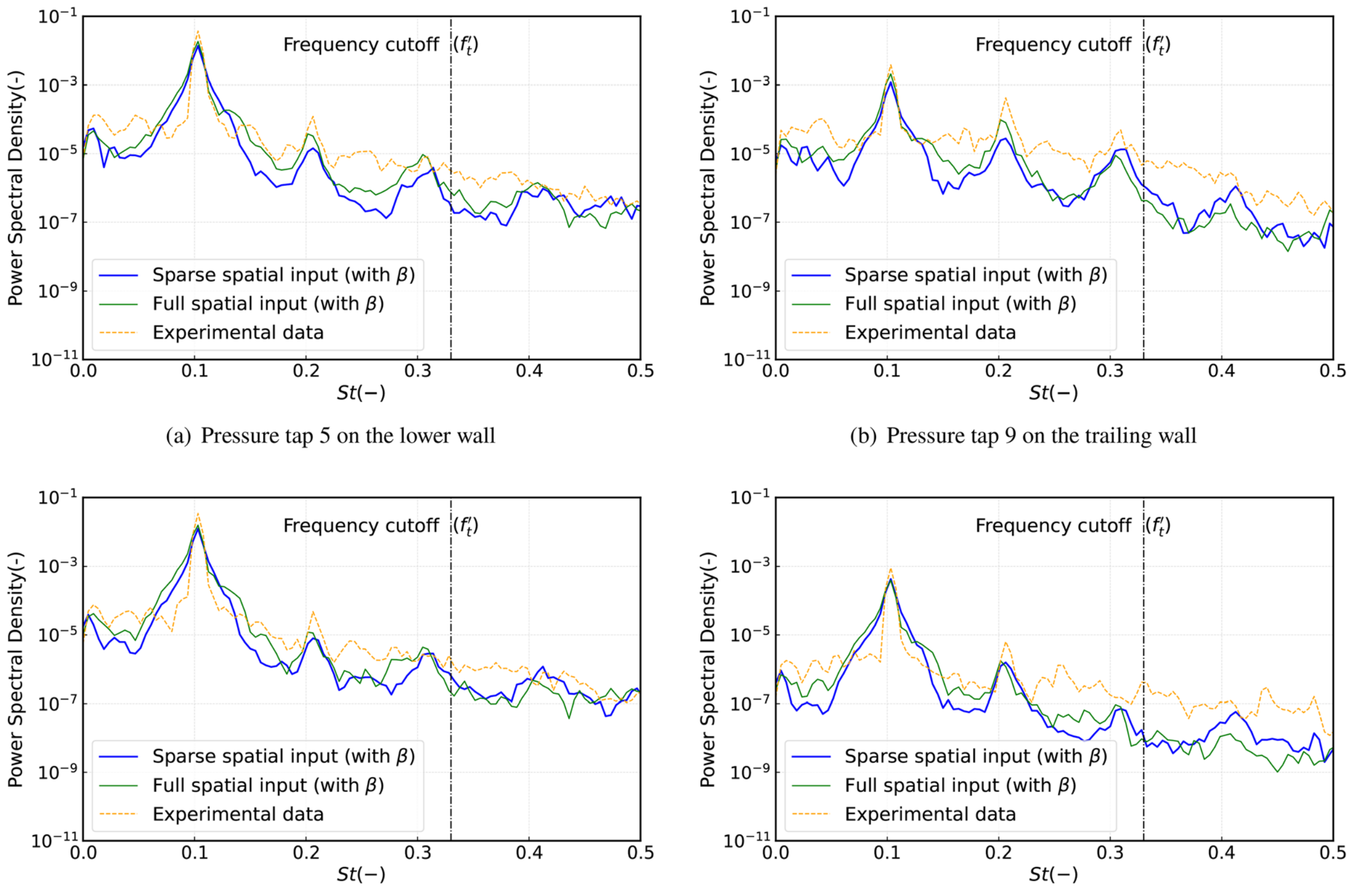

(a) Pressure tap 5 on the lower wall

(b) Pressure tap 9 on the trailing wall

(c) Pressure tap 18 on the upper wall

(d) Pressure tap 22 on the leading wall

**FIG. 20.** Forecast wall pressure fluctuating information power spectrum density (PSD) comparison at different spatial pressure taps with different spatial dimension input. In all four figures, in both x and y axis, the sign (-) denotes dimensionless value. *St* denotes the Calculated Strouhal number based on Eq. (24).

## VII. MODEL GENERALIZATION TEST

The above results and discussions in Sec. V have demonstrated the performance of the model in forecasting the wall pressure for the rectangular cylinder with side ratio 1.5, as well as the model extrapolation capability. As mentioned in Sec. IV, we have collected many datasets for a rectangular cylinder with different side ratios in the wind tunnel. In this section, we test the model generalizability in forecasting spatiotemporal wall pressure information for other rectangular cylinders with diverse side ratios. Figure 22 presents the model generalization test on the rectangular cylinders of side ratio 4 and 3.75. These two cases have distinct flow patterns with rectangular cylinder of side ratio 1.5, including flow re-attachment, vortex shedding and leading edge flow separation.[15,16,20] Therefore, the pressure distribution features of these two rectangular cylinders are distinct from the pressure distributions on the rectangular cylinder of side ratio 1.5. Figures 22(a) and 22(c) show the time-averaged wall pressure information ($\overline{C}_p$) and the standard deviation value ($\sigma$) for each pressure tap on the rectangular cylinder walls of side ratios 4 and 3.75, respectively. Figures 22(b) and 22(d) present the instantaneous wall pressure at snapshot $\Delta T/\delta t = 983$ for the rectangular cylinder of side ratios 4 and 3.75, respectively. From both time-averaged pressure ($\overline{C}_p$), standard deviation value of the pressure information ($\sigma$) and instantaneous wall pressure distributions, we can derive the model has generalization ability on unseen cases.

## VIII. CONCLUSIONS AND FUTURE PLANS

In this work, the DeepUFNet model is developed to forecast the spatiotemporal wall pressure of the rectangular cylinder. The DeepUFNet model consists of a UNet structure and a Fourier neural network. Frequency truncation is imposed in DeepUFNet, and high-frequency fluctuation is approximated with the Fourier neural network. In the UNet structure, the residual connection and batch norm are used to improve the model performance. In particular, during the DeepUFNet model training, the physical frequency loss control coefficient $\beta$ is embedded to optimize the model performance in the high-frequency regime. The $\beta$ is the dynamic coefficient depending on the progress of the training epoch number and adjusts the influential factor of physical loss during the training of the DeepUFNet model. To evaluate the model performance, the forecast result is evaluated and compared to the experimental measurements from statistical and physical interpretation on the rectangular cylinder of side ratio 1.5, and generalization ability is tested on other rectangular cylinder cases.

From the statistical comparison between the forecast result and the experimental measurements, it can be derived that the DeepUFNet model can forecast the temporally averaged pressure coefficient information, while the standard variation values vary. Embedding the physical high-frequency loss control coefficient $\beta$ in the total loss design can enhance the capacity of the model to forecast the standard deviation value and eventually improve the forecast of the spatial-temporal wall pressure. In addition, the comparison of probability density function (PDF) reveals that the DeepUFNet model can find the pressure coefficient $C_p$ that leads PDF peaks, and the forecast result has a higher PDF value in the peaks than the peaks computed from the experimental data. The fluctuating pressure in the temporal domain is investigated at different spatial points, and it is found that the incorporation of the physical high-frequency loss control coefficient $\beta$ can improve the performance of the DeepUFNet model. The frequency truncation in the high-frequency part can enhance model ability to forecast both low-frequency and high-frequency fluctuations. This finding suggests that embedding some prior physical information in the DeepUFNet model can optimize the performance not only locally, but globally in the frequency domain.

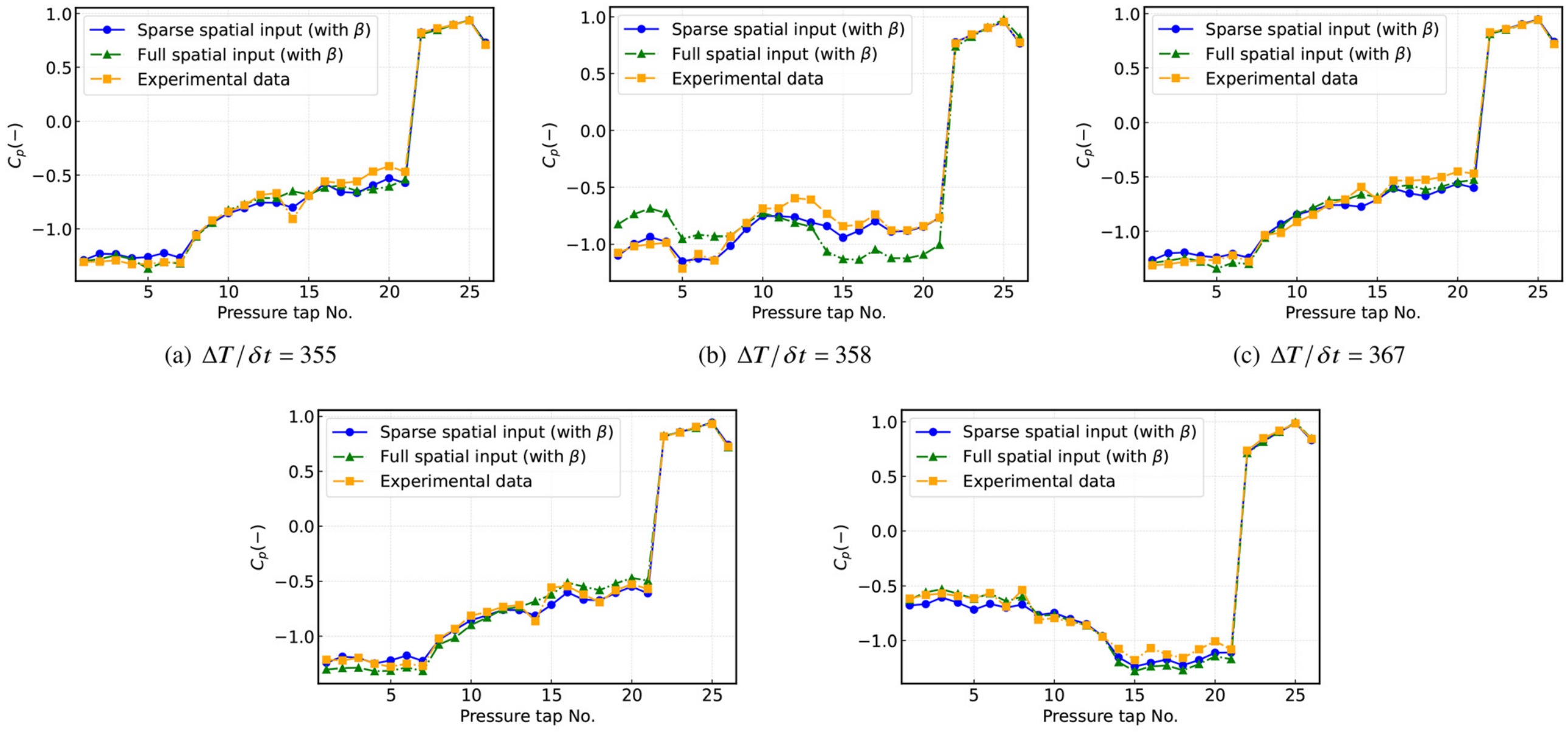

(a) $\Delta T/\delta t = 355$ (b) $\Delta T/\delta t = 358$ (c) $\Delta T/\delta t = 367$

(d) $\Delta T/\delta t = 983$ (e) $\Delta T/\delta t = 989$

**FIG. 21.** Three types of spatial pressure ($C_p$) distribution in different temporal instantaneous snapshots, forecast wall pressure with sparse spatial input (with $\beta$), forecast wall pressure with full spatial information (with $\beta$), and experimental data. In y axis, the sign (-) denotes dimensionless value of pressure coefficient. In x axis, $\Delta T$ denotes the physical time information accumulating from the start of forecast with unit *s*, $\delta t$ denotes the pressure scanning time interval with unit *s*.

The spatial pressure distribution at selected instantaneous snapshots is compared with the experimental data. It is found that when the lift coefficient lies in peaks and valleys, the forecast result aligns with the experimental data at most pressure measurement taps. The spatiotemporal correlation of the pressure on the upper wall is calculated for both the forecast result and the experimental data. The quantitative analysis shows that the DeepUFNet model gives higher spatiotemporal correlation values on the rectangular cylinder upper wall compared with experimental measurements.

The DeepUFNet model is further tested with extrapolation ability by reducing the input to sparse spatial wall pressure information. Three main investigations, including statistical information, power spectrum density of fluctuating pressure, and spatial distribution, have indicated that the DeepUFNet model can perform a satisfactory extrapolation. The extrapolation ability test reflects the robustness and generalization of the DeepUFNet model when reducing the input spatial locations on the upper and lower walls. Finally, the DeepUFNet is tested on generalization capability on other rectangular cylinders. The general test results present alignments compared to the experimental data.

In the future, there are several aspects to further advance this work. Although spatiotemporal wall pressure is forecast and studied from both spatial and temporal perspectives, our current forecast horizon of 1000 snapshots (approximately 2.5 s in physical time) remains relatively short. Extending the temporal prediction would make the model more applicable to practical engineering problems. Moreover, while the present validation is performed using data from uniform inflow past a two-dimensional rectangular cylinder, more realistic datasets can be explored, such as flows with different turbulence intensities, three-dimensional configurations, and more complex geometries. Another promising direction is to evaluate model robustness under different frequency truncations in the Fourier neural network and the physics-based high-frequency loss component.

Furthermore, the deep learning model proposed in this study has broad prospects for practical applications, especially providing robust technical support and innovative ideas for the field of wind engineering. For example, by integrating the model with real-time monitoring data from sparse sensors on real building surfaces, it is possible to achieve spatiotemporal prediction of wind pressure on building surfaces. On the one hand, it can accurately identify areas where instantaneous peak wind pressure may occur. By deploying protective measures in advance, these areas can be effectively protected, reducing the potential damage risk on the building facade.[120,121] On the other hand, based on the real-time wind pressure data forecasted by the model, decision-making support can be provided for wind load control methods, such as offering effective references for active dampers.[122,123] Overall, the deep learning model is expected to significantly enhance building safety in complex wind environments, providing new directions for future building design, wind hazard protection, and intelligent building control.

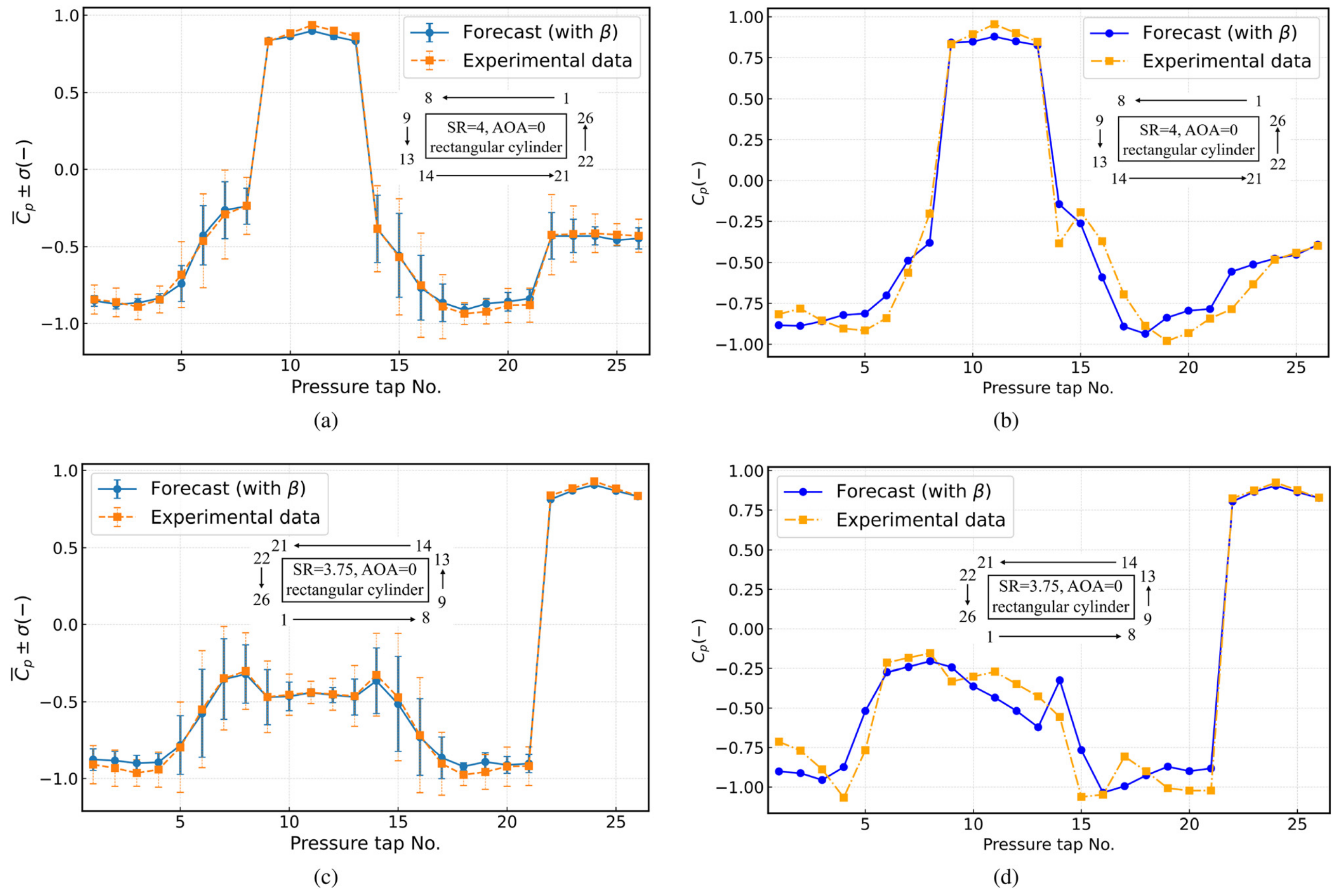

FIG. 22. Forecast result and experimental data comparison for rectangular cylinder of side ratio 4 and 3.75 under angle of attack 0. (a) Time-averaged wall pressure ($\overline{C}_p$) and standard deviation ($\sigma$) value on each pressure tap for the rectangular cylinder of side ratio 4. (b) Instantaneous wall pressure distribution on snapshot $\Delta T/\delta t = 983$ for the rectangular cylinder of side ratio 4. (c) Time-averaged wall pressure ($\overline{C}_p$) and standard deviation ($\sigma$) value on each pressure tap for the rectangular cylinder of side ratio 3.75. (d) Instantaneous wall pressure distribution on snapshot $\Delta T/\delta t = 983$ for the rectangular cylinder of side ratio 3.75. In all four figures, the sign (-) in y axis denotes the dimensionless value.

## ACKNOWLEDGMENTS

J.L. and K.T. acknowledge support from Hong Kong Research Grants Council (RGC) General Research Fund (Project No. 16211821). G.H. acknowledges support from National Natural Science Foundation of China (Project No. 52441803), Aero Science Foundation of China (2024M034077001), and Shenzhen Science and Technology Program (KJZD20230923115210021, GXWD20231129111527001).

## AUTHOR DECLARATIONS

### Conflict of Interest

The authors have no conflicts to disclose.

### Author Contributions

**Junle Liu:** Conceptualization (equal); Data curation (equal); Formal analysis (equal); Methodology (equal); Visualization (equal); Writing – original draft (equal); Writing – review & editing (equal). **Chang Liu:** Formal analysis (equal); Methodology (equal); Validation (equal); Writing – review & editing (equal). **Yanyu Ke:** Methodology (equal); Visualization (equal); Writing – review & editing (equal). **Wenliang Chen:** Visualization (equal); Writing – review & editing (equal). **Kihing Shum:** Validation (equal); Visualization (equal); Writing – review & editing (equal). **Tim K. T. Tse:** Funding acquisition (equal); Methodology (equal); Project administration (equal); Supervision (equal); Writing – review & editing (equal). **Gang Hu:** Funding acquisition (equal); Investigation (equal); Project administration (equal); Supervision (equal); Validation (equal); Writing – review & editing (equal).

## DATA AVAILABILITY

The data that support the findings of this study are openly available in GitHub at https://github.com/LIUJUNLE97/DeepUFNet

(Ref. 125) and in Zenodo at https://doi.org/10.5281/zenodo.16739156 (Ref. 124).